\documentclass[10pt,a4paper]{report}
\pdfoutput=1

\usepackage[utf8]{inputenc}
\usepackage{booktabs}
\usepackage{graphicx}

\usepackage{multirow}
\usepackage{listings}

\newcommand{\HRule}{\rule{\linewidth}{0.5mm}}

\newcommand\circlenum[1]{\textcircled{\scriptsize #1}}

\usepackage{amsmath}

\usepackage{graphicx}
\usepackage{subcaption}

\usepackage{caption} 
\usepackage{pgfplots}
\pgfplotsset{compat=newest}
\usetikzlibrary{calc}
\makeatletter
\let\percent\@percentchar
\makeatother

\usepackage[left=3.2cm, right=3.2cm]{geometry}

\usepackage{titlesec}
\titleformat{\chapter}{\bfseries\Huge}{\thechapter}{0.7cm}{}

\begin{document}

% % % % % % % % % % % % % % % % %
% TITLE PAGE

\begin{titlepage}

	\begin{center}
	
		% inst. logo
		\includegraphics[width=0.4\linewidth]{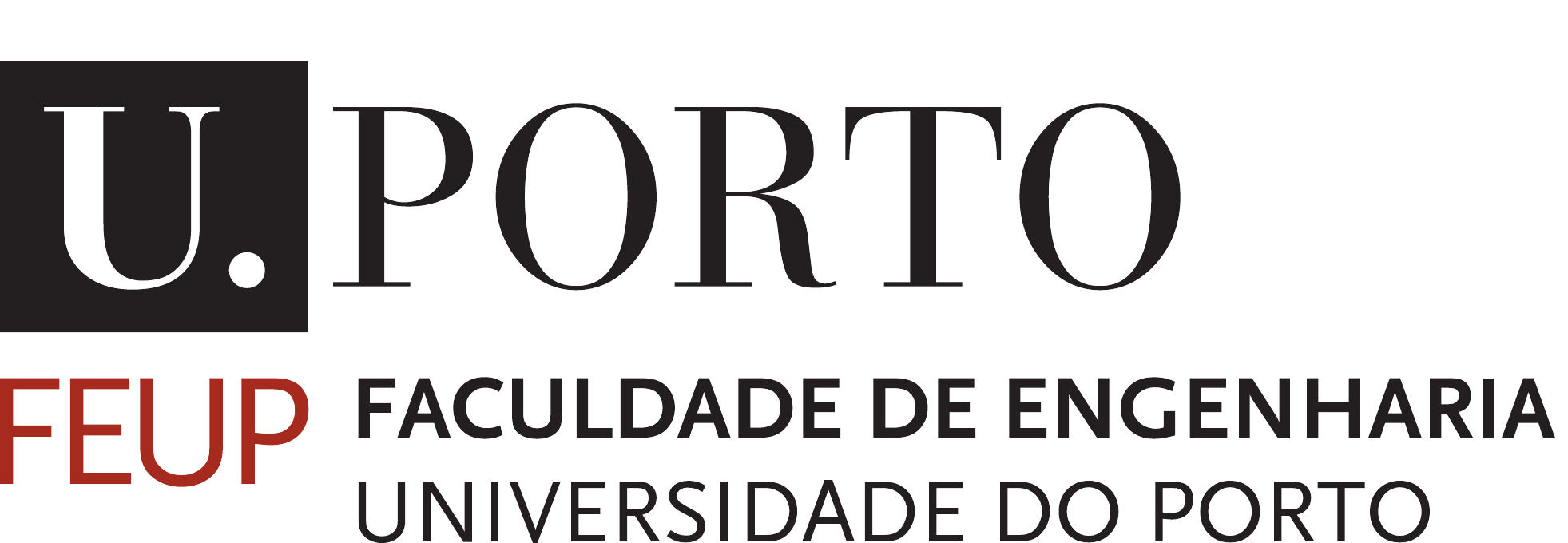}~
		\\[1cm]
%		\hspace*{4.2cm}~%

		% inst. name
		\textsc{{\Large Department of Informatics Engineering \\[-0.15cm]} {\LARGE Faculty of Engineering, \\[0.15cm] University of Porto}}\\[1.5cm]
	
		\vfill

		\includegraphics[width=0.4\linewidth]{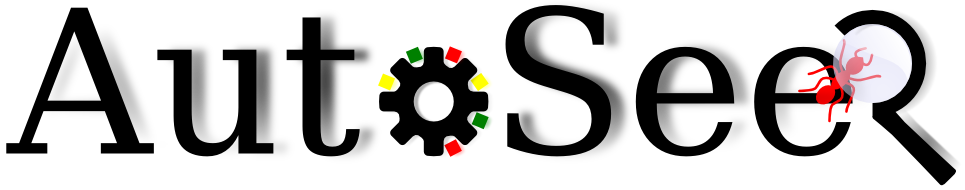}
	
		% document type
		\textsc{\Large Technical Report}\\[0.5cm]
	
		% title
		\HRule \\[0.2cm]
		{\huge \bfseries Fault Detection in C Programs using Monitoring of Range Values \\[0.2cm]}
		{\LARGE \bfseries --- Preliminary Results --- \\[0.2cm]}
		\HRule \\[1.5cm]
	
		% authors
		{\Large Pedro Pinto \\ \texttt{p.pinto@fe.up.pt} \\[0.3cm]}
		{\Large Rui Abreu \\ \texttt{rui@computer.org} \\[0.3cm]}
		{\Large João M. P. Cardoso \\ \texttt{jmpc@fe.up.pt}}
	
		% date
		\vfill
		{\Large February 1, 2015}
	\end{center}

\end{titlepage}

% % % % % % % % % % % % % % % % %
% ACK AND ABSTRACT

% Acknowledgments
\renewcommand{\abstractname}{Acknowledgments}
\begin{abstract}

This work was partially funded by the ERDF through the Programme COMPETE and by the Portuguese Government through FCT - Foundation for Science and Technology, project reference FCOMP-01-0124-FEDER-020484.

The authors are grateful to the project partners, namely Critical Software, which gently provided the source code of one of the example applications used to evaluate this work.

For aspects related to the LARA language, its weaver, and the prompt assistance on technical aspects of the work through its several phases, the authors would like to thank Tiago Carvalho.

Pedro Pinto would also like to acknowledge the support of the Doctoral Program in Informatics Engineering, in the form of a first year PhD scholarship, without which, his enrollment in the doctoral program would have not been possible.
\end{abstract}

% Abstract
\renewcommand{\abstractname}{Abstract}
\begin{abstract}
% part of the autoseer project, what is autoseer
This technical report presents the work done as part of the AutoSeer project, which investigated the use of various generic invariants in the value and time domain, their effect on Spectrum-based Fault Localization's diagnostic precision, their relation with existing test oracles, and their runtime overhead, in particular, the density required or strategic placement (trading off overhead versus precision).

Our work in this project was to develop a source-to-source compiler, MANET, for the C language that could be used for instrumentation of critical parts of applications under testing. The intention was to guide the compilation flow and define instrumentation strategies using the Aspect-Oriented Approach provided by LARA\footnotemark. This allows a separation of the original target application and the instrumentation secondary concerns.

One of the goals of this work was the development of a source-to-source C compiler that modifies code according to an input strategy. These modifications could provide code transformations that target performance and instrumentation for debugging, but in this work they are used to inject code that collects information about the values that certain variables take during runtime. This compiler is supported by an AOP approach that enables the definition of instrumentation strategies. We decided to extend an existing source-to-source compiler, Cetus, and couple it with LARA, an AOP language that is partially abstracted from the target programming language.

We propose and evaluate an approach to detect faults in C programs by monitoring the range values of variables. We consider various monitoring strategies and use two real-life applications, the GZIP file compressor and ABS, a program provided by an industrial partner. The different strategies were specified in LARA and automatically applied using MANET. The experimental results show that our approach has potential but is hindered by not accounting for values in arrays and control variables. We achieve prediction accuracies of around 54\% for ABS and 83\% for GZIP, when comparing our approach to a more traditional one, where the outputs are compared to an expected result.

\footnotetext{LARA~\cite{cardoso2012lara, Cardoso2014, y} is an Aspect-Oriented Programming language (AOP) specially designed for allowing developers to program code instrumentation strategies, to control the application of code transformations and compiler optimizations, and to effectively control different tools in a toolchain. LARA provides a separation of concerns, including non-functional requirements and strategies, for the mapping of high-level applications to high-performance heterogeneous embedded systems. Moreover, LARA provides developers with artefacts and with a unified view which diminishes the efforts to program and evaluate Design Space Exploration (DSE) schemes. The research and development of the LARA language in the context of the FP7 REFLECT research project was strongly driven by industry requirements (e.g., Honeywell, Coreworks, and ACE).}

\end{abstract}

% % % % % % % % % % % % % % % % %
% INTRODUCTION

\chapter{Introduction}

% part of the autoseer project, what is autoseer
This report presents the work done as part of the AutoSeer\footnote{For more information on the AutoSeer project, please refer to: http://autoseer.fe.up.pt/} project. AutoSeer investigated the use of various generic invariants in the value and time domain, their effect on Spectrum-based Fault Localization's diagnostic precision, their relation with existing test oracles, and their runtime overhead, in particular, the density required or strategic placement (trading off overhead versus precision).

% what is our work inside autoseer
More specifically, our work in this project was to develop a source-to-source compiler, MANET, for the C language that could be used for instrumentation of critical parts of applications under testing. The intention was to guide the compilation flow and define instrumentation strategies using the Aspect-Oriented Approach (AOP)~\cite{irwin1997aspect} provided by LARA~\cite{cardoso2012lara, Cardoso2014}. This, among other advantages, allows a clear separation of the original target application and the secondary concerns represented by our source code instrumentation.

% what were our goals, first goal
There were two main goals for this work. First, the development of a compiler that would take applications written in the C language and output modified C code according to an input strategy. In theory, these modifications could provide code transformations that target performance and instrumentation for debugging. In this work, this compiler is used to inject code that collects information about the values that certain variables take during runtime. This compiler  supported by an AOP approach that enables the definition of instrumentation strategies, for instance, in our case, deciding which variables to monitor and how to collect and use this information. We decided to extend an existing source-to-source compiler, Cetus~\cite{cetus}, and couple it with LARA~\cite{cardoso2012lara, Cardoso2014}, an AOP language that is partially abstracted from the target programming language.

% why is a LARA-controlled C2C framework important? what would be the alternatives?
A source-to-source compiler framework controlled by LARA strategies provides important advantages over code instrumentation alternatives. Previous uses of LARA, namely in the context of source-to-source transformations and instrumentations strategies (see, e.g., the version of the Harmonic~\cite{x} source-to-source framework controlled by LARA~\cite{y}), have already highlighted and exposed strong evidence of its advantages~\cite{Cardoso2014}. There are possibilities such as manually instrumenting the source code but that has obvious downsides as it is time-costly and error-prone. Other approaches include binary instrumentation which can be less flexible and also more complex. There is also the possibility of extending existing compilers. However, per each instrumentation strategy we wanted to apply, a new compiler pass would have to be written. This is usually out of the scope and skills of users that need to instrument applications. All of these alternative approaches hinder the possibility of exploring different strategies, which is of utmost importance, especially in the early stages of a project.

% second goal
The second goal was the development and the evaluation of instrumentation strategies that could be used to detect faults present in a given application. Our compiler, coupled with LARA allows a great flexibility in the type of instrumentation that is possible. We can, with fine-grain control, select and act over a multitude of points of interest. For instance, it is possible to develop strategies that test program paths by instrumenting control-flow statements. For this particular work, we explored the possibility of using the range taken by certain variables during the execution, so all strategies revolve around this idea. We change the sets of variables chosen for monitoring for a particular program.

% illustrate how the range has been previously used
The computing of the range of values (known as range-value analysis) has been addressed since many years~\cite{Asaithambi82}. Approaches may rely in static analysis, monitoring, and a mix of both. Range-value analysis has been used in the context of compiler and hardware synthesis optimizations for reducing the bitwidths of variables, see, e.g, a static approach~\cite{Stephenson00}, and, e.g., a monitoring approach using simulation~\cite{Brooks99}. Research efforts also consider range-value analysis in the context of detecting program faults. For instance, in the context of fault screeners, the range of a variable is used to identify collar variables~\cite{Santos12}. They consider that if the range is constant across executions, then the variable should not monitored. With this approach they reduced by about 50\% the number of variables to monitor for the 3 applications considered. Research efforts in the context of security vulnerabilities, have also used range value analysis to detect program faults (e.g., buffer overflows~\cite{Simon08, Wagner00}). 

% other fault screener approaches
Using program invariants as mean to locate faults is already a proven approach~\cite{4,9}. Although we decided to use ranges taken by variables, there are other approaches that focus on using other types of program invariants, which may prove more beneficial in different contexts. Daikon~\cite{10} is a platform for dynamic invariant detection. It runs a program and finds likely invariants such as constants and ranges. It supports multiple programming languages and can be extended by the user to detect new invariant types. Zoltar~\cite{13} is a toolset for automatic fault localization that instruments application-specific fault screeners. Zoltar provides, by default two Spectrum-based Fault Location techniques but can be extended easily. There are also approached that work on the hardware level. IODINE~\cite{12} is a tool that captures properties from hardware design simulations that can be used as dynamic invariants. These properties include state machine protocols and request-acknowledge pairs. iSWAT~\cite{15} is a framework for detection of permanent faults at the hardware level that aims to reduce the latency and increase the coverage of previous techniques. Compile-time instrumentation is used to monitor values during training and generate the invariants. Code insertion is then used to inject the invariants used at runtime.s

% why use variables? pros and cons
We decided to monitor variables with the purpose of detecting faults as they are integral part of the computation of any application and, as such, we can expect faults to eventually propagate to a variable. It is also straightforward to monitor them, in the sense that it is easy to locate the sites in the code where they are updated. This however, brings the obvious downside that we will, by default, monitor a large number of places. Therefore, choosing the most critical variables becomes the main issue, as the so called \textit{collar variables}~\cite{5} may not be easily identifiable. Finally, it is simple to monitor the range of values that a variable takes during execution, as we need to maintain only, for each monitored variable, a the minimum and the maximum values. This is not to say that for a large number of variables, this will not be computationally expensive or require too much memory, which can be a problem in certain computing systems.

% explain out approach using the range
The approach using ranges is defined as follows. Initially, we have a training phase in which we execute the application using several test cases, trying to exercise all of its modules and features. When we do this, and using our instrumentation, we learn a range for each variable, in each function. This is basically the minimum and maximum values that the variable took during execution in the training phase. Then, during the execution or testing phase, we execute the same application using similar instrumentation to detect if any of the learned ranges was violated. That is, if a given variable took a value that was either lesser than the minimum or greater than the maximum learned in the training phase. This detection could also be performed online, on a deployed application, which would trigger whatever fault tolerance measures were in place. For our work, however, this range violation is tested offline as we only intend to measure the diagnostic accuracy of this technique.

% how did we evaluate our approach, what applications were used
In order to evaluate this approach we used two different applications. First, we used one of the original versions of the \textit{GZIP}\footnote{The application source code and test cases for GZIP were provided by the Software-artifact Infrastructure Repository: http://sir.unl.edu/portal/bios/tcas.php.} application, which is found in Linux distributions. This program can be used to compress and decompress files which are then typically archived. The second application is a simulation of an Anti-lock Breaking System (ABS)\footnote{The application source code for ABS was gently provided by our industrial partner, Critical Software: www.criticalsoftware.com.} that will, given the initial car and wheel speeds, calculate the distance needed to stop the car.

% report structure
The remainder of this report is structured as follows. Chapter \ref{chap:compiler} briefly presents the LARA language and details the architecture of MANET, our source-to-source compiler that is supported by LARA. In Chapter \ref{chap:strats}, we explain our approach to detect application faults using range values, as well as the monitoring strategies developed and evaluated in this work. The description and results of the conducted experiments are shown in Chapter \ref{chap:exp}. Finally, Chapter \ref{chap:conc} presents the conclusions drawn from this work as well as some possible paths for future research.

% % % % % % % % % % % % % % % % %
% COMPILER

\chapter{Source-to-Source Compiler}\label{chap:compiler}

In this chapter, we briefly present the LARA language and it is used with our compiler, MANET. Then, we detail the architecture of MANET and the compilation flow that is used to weave a LARA strategy with an application written in C. We also present some of the transformation engines that are used in this work and illustrate how the framework can be used with simple examples.

\section{The LARA Language}
Although all programming paradigms have their decomposing criteria (e.g., function decomposition or objection-oriented decomposition), there are concerns that do not align with the primary decomposition~\cite{irwin1997aspect}. These secondary or cross-cutting concerns, are spread along the program decomposition unit, usually repeated and scattered along the source code, polluting the primary objective of the system~\cite{elrad2001aspect}.

Aspect Oriented Programming (AOP)~\cite{irwin1997aspect} is a paradigm which intends to increase the modularity of programs, by separating cross-cutting concerns from those that describe the intended behavior of the application~\cite{elrad2001aspect}. This paradigm is based on the idea that a system works better if its properties, non-functional requirements and areas of interest are specified separately and there is an \textit{aspect} describing the relationships between them. The \textit{weaving} process will then bring these concerns together into the intended program~\cite{elrad2001aspect}. Using aspects to separate secondary concerns from the core objective of the program results in cleaner code, easier concern analysis, monitoring, tracing and debugging.

Currently, there is a diverse set of aspect programming languages, such as AspectJ~\cite{kiczales2001overview}, an aspect-oriented extension for Java, and AspectC++~\cite{spinczyk2002aspectc++}, to apply aspects in C++ programs. 

% LARA

The previously mentioned AOP languages define specific aspects, which cannot be reused across different aspect and target languages. For instance, an extensive aspect defined in AspectJ, possibly useful for Java, cannot be immediately used with other languages without manual translation. Other disadvantages of the mentioned tools include their simple model for points of interest (that, e.g., does not consider loops and local variables) and the reduced number of available actions, usually restricted to code insertion.

LARA~\cite{cardoso2012lara, Cardoso2014} is an AOP approach that is partially agnostic to the target programming language. It allows the definition of aspects for different target languages, i.e., LARA aspects are generic enough so they can be applied to several problems, over a number of (specified) languages. LARA enables developers to capture non-functional requirements and concerns in the form of strategies, which are completely decoupled from the functional description of the application~\cite{cardoso2012lara,cardoso2013lara}.

LARA differs from other AOP approaches as its syntax is agnostic to the contents of the target language. Instead it relies on a external language definition which contains three generic components, namely, join points, attributes and actions. The \textit{join points}, are points of interest in the code that one intends to observe or influence, e.g., function calls or array accesses. \textit{Attributes}, represent join point information that we can retrieve, for instance, the name of the called function. Activities one can carry out over the selected join point, such as monitoring code insertion, are called \textit{actions}. LARA provides access to several types of actions, compared to current state-of-art approaches, that usually focus on code injection (e.g., \cite{kiczales2001overview} and \cite{spinczyk2002aspectc++}). Besides allowing code insertion, LARA enables attribute definition of a specified join point to a new intended value, such as variable type redefinition and function renaming. Moreover, additional types of available actions may be defined in the language specification, such as optimizing tasks, software/hardware partitioning, and code transformations.

With LARA, the language specification is defined externally, as opposed to other common approaches. This allows flexibility for new updates and to easily target different programming languages. As a consequence, LARA does not have a built-in weaving process. Instead, the LARA Compiler parses the aspects and combines them with the language specification to generate the Aspect-IR, an intermediate representation of the aspects for a specific target language. The Aspect-IR should then be interpreted by an external interpreter or weaving tool, such as LARAI~\cite{cardoso2013lara}, the LARA Interpreter. LARA remains partially language-agnostic, as it is the weaving tool that is bound to a specific target language.

A concern intended to be applied over the target application, is expressed as an aspect definition, or \textit{aspectdef}, the basic modular unit of LARA. An aspect is comprised of three main steps, which will carry out the intended concern. First, we need to capture the points of interest in the code using a \textit{select} statement. Now, using the \textit{apply} statement, we are able to act over the select points, if certain concern constrains are met. These represent the last required step, and can be defined as a filter inside the \textit{select} statement for simple constrains, or using a \textit{condition} statement if they are more complex. To allow the definition of more sophisticated concerns, it is possible to embed JavaScript code inside LARA aspects.

\begin{figure}[h]
	\begin{center}
\begin{lstlisting}[boxpos=b]
aspectdef LoopIterations

 var id = 0;
 select loop.first_stmt end
 apply
 
  $loop.insert before 'int counter_[[id]] = 0;';
  $loop.insert after
   'printf("loop [[$loop.rank]]:%d\n", counter_[[id]]);';
  $first_stmt.insert before 'counter_[[id]]++;';
  id++;
 end
 condition $loop.type == 'for' end
end
\end{lstlisting}
	\end{center}
	\caption{Simple LARA aspect that counts and prints the number of iterations of a loop.}
	\label{fig:loop_iterations_runtime}
\end{figure}

Figure \ref{fig:loop_iterations_runtime} shows an example aspect written in LARA that inserts code around loops to count the number of iterations at runtime. The aspect starts by selecting loops and their first statement (line 3). Then, we insert code before the loop to initialize a counter variable, unique to the loop. We insert a call to print the number of iterations after the loop. Finally, we add an increment statement at the beginning of the loop body, i.e., before the first statement. These code insertion actions appear in lines 5 through 12, and are only applied if the condition in line 13, which states the loop must be of type \textit{FOR}, is met.

\section{MANET: Extending Cetus with LARA}
Our solution is MANET, a source-to-source compiler for C, that is controlled through an AOP approach, using LARA aspects. This compiler manages to leverage the expressiveness and modularity of LARA aspects, to control the query and manipulation of an Abstract Syntax Tree (AST) of an existing compiler infrastructure, Cetus~\cite{cetus}. This creates an easy compilation process of C source files with the main goal of code instrumentation. The usage of aspects enables an easy selection of points of interest in the code, represented by LARA join points, which can then be analyzed for information retrieval or transformed through actions. Thus, MANET can be used to create information reports based on compiler analyses or to implement complex and sophisticated code instrumentation and transformation strategies. MANET makes use of two already existing and tested platforms, LARA, the aspect-oriented DSL used to control the compilation flow, and Cetus used as both intermediate representation and back-end component. We believe that using two already proved and tested platforms brings value and robustness to the solution.

As a source-to-source compiler, MANET is intended to be used as an intermediate tool, part of a more complex toolchain, rather than as a standalone compiler. This usage is perfectly illustrated in \cite{dse}, where MANET is used to transform source code as part of a Design Space Exploration mechanism that aims to increase performance on specific kernels using different multicore models. Despite being in its early stages of development, we believe MANET is a flexible source-to-source compiler suitable for a variety of tasks, such as code instrumentation, code transformations and optimizations and even as an alternative to multi-architecture, directive-based compilation.

The MANET compilation flow is guided by LARA aspects that describe the intended strategies. As an example, consider the aspects depicted in Figure \ref{fig:lara_examples}. In \textit{PrintLoopIterations}, we print to the console the number of iterations each loop is going to execute. The loops are identified by their \textit{rank} attribute, which uniquely identifies a loop inside a source file. This aspect shows the most basic LARA construct, a \textit{select} followed by an \textit{apply}. The former is used to capture points of interest in the source code, loops in this example, and the later is used to act over these points. The \textit{InstrumentFunctionCalls} aspect inserts monitoring code (line 13) that will print a message with the called functions and their caller. Finally, there is an example of a LARA-controlled loop transformation in the aspect \textit{UnrollLoopsBy2}. This aspect captures innermost \textit{FOR} loops and transforms them with Loop Unrolling.

\begin{figure}[h]
	\begin{center}
\begin{lstlisting}[boxpos=b]
aspectdef PrintLoopIterations
 select loop end
 apply
  println($loop.rank + ":" + $loop.num_iterations);
 end
end

aspectdef InstrumentFunctionCalls
 select function.call end
 apply
  var func = $function.name;
  var fcall = $call.name;
  insert after 'printf("[[func]]->[[fcall]]");';
 end
end

aspectdef UnrollLoopsBy2
 input funcName end
 select function{name==funcName}.loop{type=="for"} end
 apply
  perform Unroll(2);
 end
 condition $loop.is_innermost end
end
\end{lstlisting}
	\end{center}
	\caption{Simple examples of LARA aspects that illustrate how MANET performs different tasks, such as code instrumentation or transformation.}
	\label{fig:lara_examples}
\end{figure}

%%%%%%%%%%%%%%%%%%%%%%%%%%%%%%%%%%
%%%%%%  MANET Architecture  %%%%%%
%%%%%%%%%%%%%%%%%%%%%%%%%%%%%%%%%%
\subsection{MANET Architecture}

\begin{figure}[h]
	\begin{center}
		\includegraphics[width=\columnwidth]{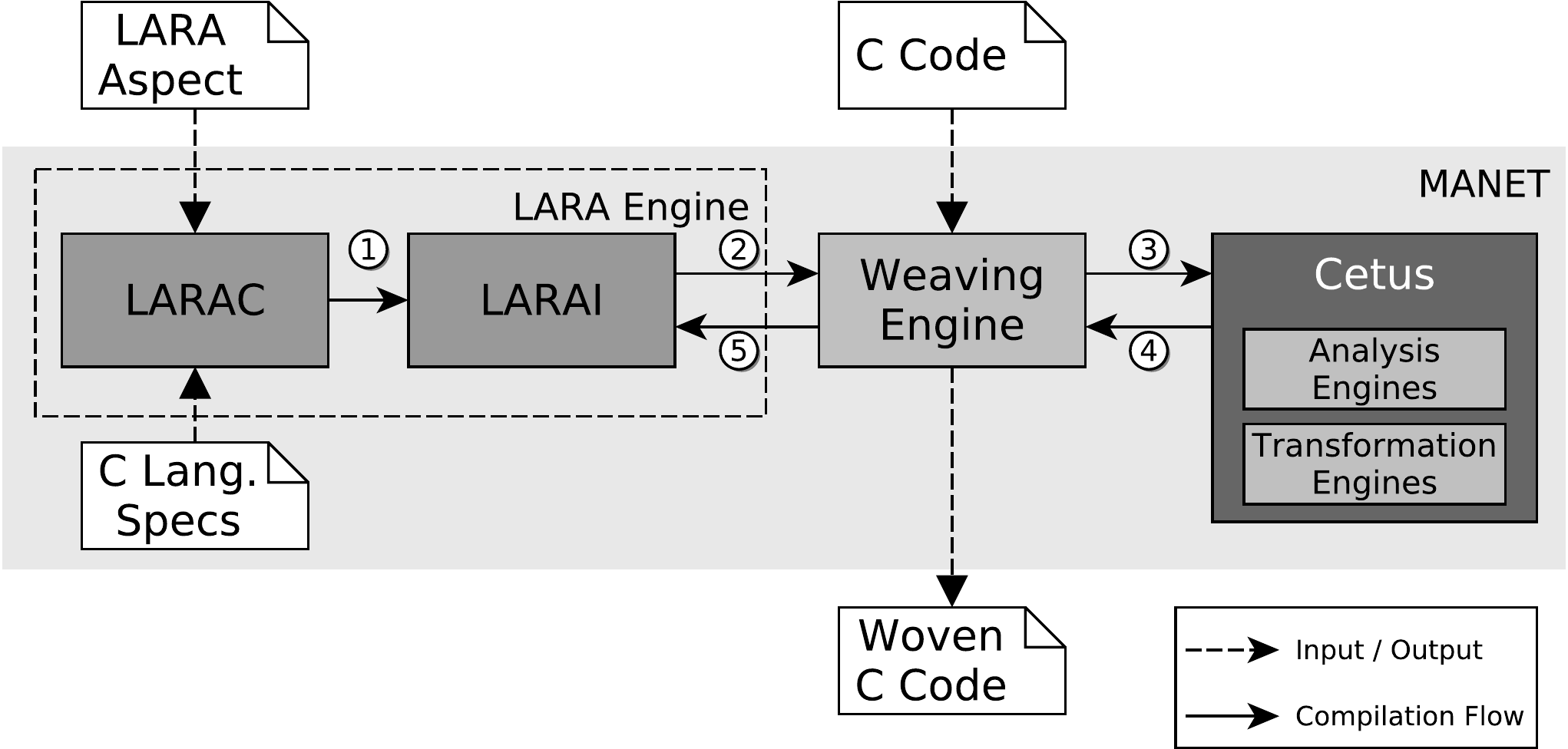}
	\end{center}
	\caption{The architecture of MANET.}
	\label{fig:MANET_arch}
\end{figure}

MANET is composed of three main components, which are presented in Figure \ref{fig:MANET_arch}. The first component is the LARA Engine, that provides the interface to the user and translates the aspects into commands for the next component. The Weaving Engine is responsible for establishing communication between the other two components. The commands from the LARA Engine are converted into specific actions that are applied over Cetus, which builds and maintains the AST after parsing the input C source files. Each of these components is detailed below.

\subsubsection{LARA Engine}
The LARA Engine is the component where the execution starts and is itself composed of two subcomponents. The first, is the LARA Compiler, which is responsible for compiling the aspect into an intermediate representation, known as Aspect-IR. Because LARA is partially language-agnostic, it also needs as input a description of the target language. This is the \textit{C Lang. Specs} block in Figure \ref{fig:MANET_arch}, a set of three XML files (part of MANET) that contain the specification of the language: join points, attributes and actions. The generated Aspect-IR is saved to an XML file and passed to the LARA Interpreter, the second subcomponent. This is step \circlenum{1} in Figure \ref{fig:MANET_arch}.

In step \circlenum{2} of the compilation flow, the interpreter takes this new representation and communicates with the Weaving Engine the instructions that it needs to carry, i.e, join point selections, attribute queries or program-altering actions. There is an iterative information flow that goes from the LARA Interpreter, to the Weaving Engine \circlenum{2}, to Cetus \circlenum{3} and then back again to the interpreter through the reverse path, \circlenum{4} and \circlenum{5}. The information flows forward in the form of commands to the Weaving Engine and flows backwards in the form of join points or attributes. This process repeats several times during a normal execution.

The LARA Engine provides the interface with the user, as it is the component responsible for parsing the LARA aspects and initiate the execution, as well as printing the information to the console, as in the aspect \textit{PrintLoopIterations} of Figure \ref{fig:lara_examples}.

\subsubsection{Cetus}

Cetus is a source-to-source compiler infrastructure for the C language, that supports the ANSI C standard and is written in Java. This is an infrastructure aimed towards research on multicore computing with a focus on auto-parallelization. Current work~\cite{cetus} tries to translate shared memory programs written in OpenMP, into other models, such as Cuda. This is supported by several transformation and analysis passes, such as Control Flow and Data Dependence Analysis.

The AST, the main element of Cetus regarding MANET, represents the C program and is composed of Java classes such as Program, Translation Unit, Procedure, Statement and Expression. These are the basic building blocks of the C hierarchy. All these elements are used as nodes of the AST and can be visited during tree traversals. During these traversals it is possible to manipulate the nodes, effectively changing the resulting C program.

Any join point that is created during the execution is based on this AST and has a reference to the specific node it represents, e.g., a CFunction has a reference to a Procedure on the AST. Whenever there is a query about a join point attribute, a method is triggered that uses the tree reference to return the requested information. Analogously, when an action is to be performed that alters the AST, a method of the join point is called that uses the tree reference to perform the desired changes. 

\subsubsection{Weaving Engine}
The Weaving Engine is the central component of MANET. It provides a bidirectional communication channel between the other two components and is responsible for instantiating the intents described in the LARA aspect by either querying or changing the AST on Cetus.

This engine has a hierarchic structure that mimics the join point tree defined in the language specification. If the join point model defines that it is possible to select loops from inside a function, then there is a class, CFunction, from which it is possible to select instances of another class, CLoop. Similarly, this structure also follows the language specification regarding the join point attributes, as each class implements getter methods for specific attributes. For instance, CLoop has a \textit{getNumIterations()} method, as the language specification defines that it is possible to consult the number of iterations of a loop join point.

The actions defined in the language specification are performed by the Weaving Engine by interacting with the AST created by Cetus. This is performed using extensions to the Cetus infrastructure, represented by the \textit{Analysis Engines} and \textit{Transformation Engines} in Figure \ref{fig:MANET_arch}. For instance, \textit{NormalizeReturn} is a transformation pass that, among other things, changes the program structure so that every function has at least one return statement. This is a transformation that exists on Cetus and was adapted to be called from a LARA aspect. Consider also \textit{UnaryExpansion}, that transforms expressions that use unary increment or decrement operators into their equivalent binary expressions. This transformation pass was written for MANET, as it facilitates code analysis and instrumentation.

\subsubsection{Join points and the AST}

\begin{figure}[h]
	\begin{center}
		\includegraphics[width=\columnwidth]{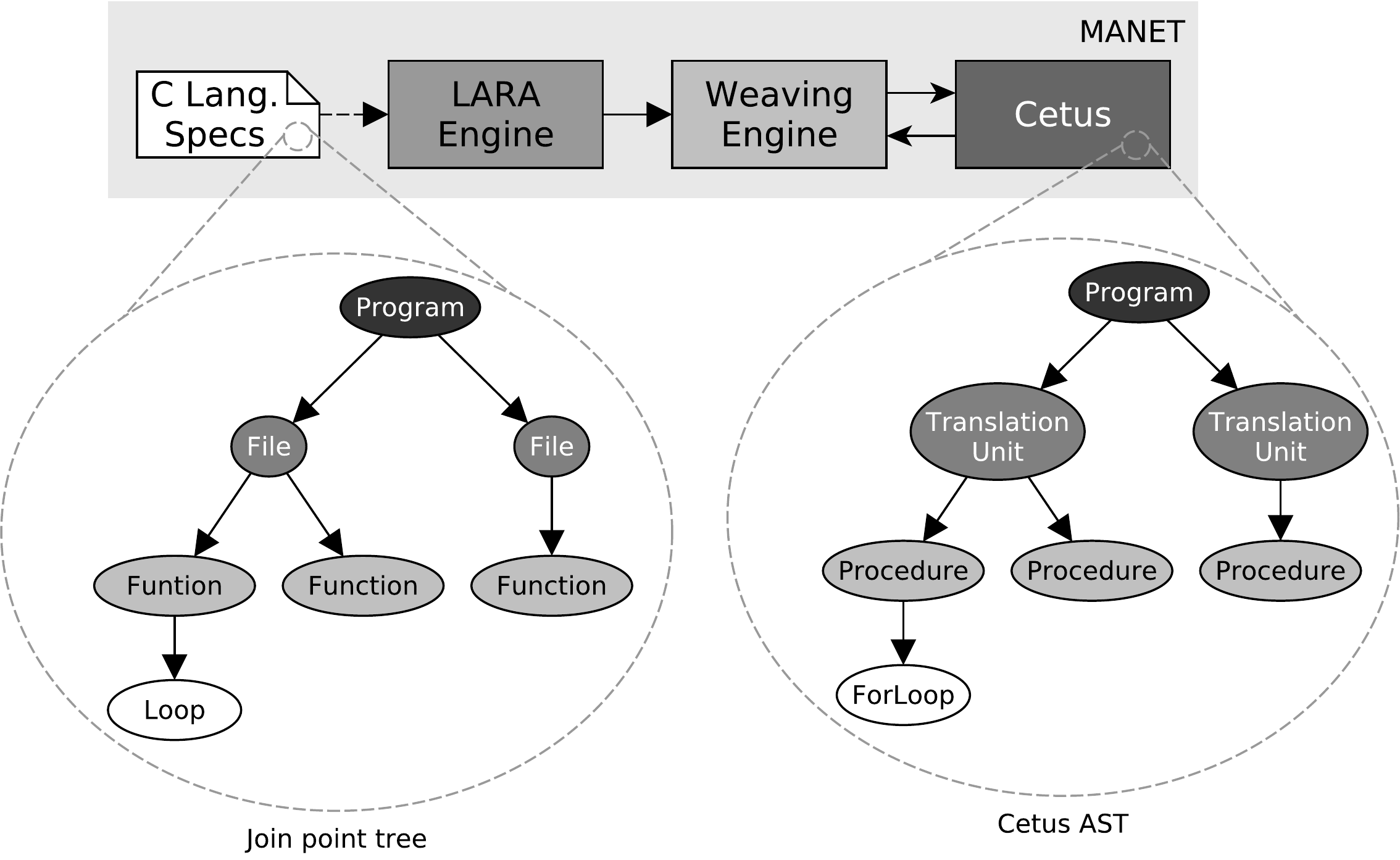}
	\end{center}
	\caption{The mapping between the join point tree and the AST created by Cetus.}
	\label{fig:MANET_arch_tree}
\end{figure}

There is a logical match between the join points used by the LARA Engine and the AST built by Cetus. Although this is not how it is actually implemented, as there is no join point tree at any time during execution, it is the most intuitive way of thinking about the relation of both representations. For instance, there is a correspondence between a TranslationUnit and a CFile join point as both represent a C source file. With rare exceptions, mainly implementation related, there is a direct, one-to-one mapping from a join point to a node on the AST. Thus, we can think of having two trees that can be translated into one another, creating a simple, yet effective and intuitive representation of the program information, as portrayed in Figure \ref{fig:MANET_arch_tree}.

In actuality, only the AST exists and the join points are extracted as needed. This is an implementation choice, taken because we want to keep the tree updated with the changes that are applied, so that any change on the tree is immediately seen by the following information queries. Consequently, it is simpler to keep a single tree, halving the updates and avoiding synchronization tasks. As expected, this choice comes with its own disadvantages, namely, having to create unnecessary join points. Picture an aspect that does not perform any change on the tree and is used only to collect information. In this case, certain join points are created repeatedly when there is no need to. Ideally, MANET would analyze the LARA aspect to check whether it contains tree-altering actions and, based on this information, select one of the two representations. This, however, might not be possible due to the highly dynamic nature of LARA.

%%%%%%%%%%%%%%%%%%%%%%%%%%%%%%%%%%%%%%
%%%%%%  Transformation Engines  %%%%%%
%%%%%%%%%%%%%%%%%%%%%%%%%%%%%%%%%%%%%%
\subsection{Transformation Engines}
MANET currently provides a number of transformation engines that can be invoked by any LARA aspect. Those engines have mainly two different goals. Some can be used to transform the code so that it is easier to analyze and instrument, e.g., ensuring that every function as one return statement, providing a well known exit point. Other engines change the code by applying well-known transformations that usually improve performance. Table \ref{tab:trans_engines} presents the current transformation engines that were used in the strategies developed for this work. The \textit{Origin} column indicates whether the transformation was already available with the Cetus distribution\footnote{More information at: http://cetus.ecn.purdue.edu/Documentation/api/} or if it was specifically developed for MANET. The transformations available in Cetus were adapted in order to be called from within LARA aspects.

\begin{table}

	{\small \begin{center}
		\begin{tabular}{ll}
			\toprule
			Engine					& Origin\\
			\midrule
			NormalizeReturn			& Cetus	\\
			SingleDeclarator			& Cetus	\\
			AssignExpansion			& MANET	\\
			StructAssignDecomposition	& MANET	\\
			UnaryExpansion			& MANET	\\
			\bottomrule
		\end{tabular}
	\end{center}}
	\caption{The transformation engines used in this work.}
	\label{tab:trans_engines}
\end{table}

\paragraph{NormalizeReturn} Guarantees that every function has at least one return statement. Additionally, it changes existing return statements so that they have their return expressions replaced with a single temporary variable, whose value is the previous return expression.

\paragraph{SingleDeclarator} If there is a declaration statement with multiple declarations, it is divided into multiple statements, each containing a single declaration.

\paragraph{AssignExpansion} Augmented assignments, such as +=, are replaced with normal assignments, replicating the left hand side in the right side.

\paragraph{StructAssignDecomposition} Replaces the right hand side expression of an assignment to a struct member with a single temporary variable. This variable contains the value of the previous right hand side expression.

\paragraph{UnaryExpansion} Expressions with unary increment and decrement operators are replaced with the equivalent assignment expressions. This only happens if the expression is, by itself,a single statement, i.e., this transformation is not performed if the unary operation is used as part of a more complex expression.

\section{Summary}

In this chapter we gave an overview of the LARA language and its main concepts (join points, attributes and actions) and its main constructs (select, apply and condition blocks). Then, we presented a detailed view of MANET, explaining how we extended the Cetus compiler and added LARA support to control the compilation tasks. There are three main components, the LARA Engine, the Weaving Engine and the extended Cetus compiler. These interact to compile the C source files according to the defined strategies. The LARA engine compiles and interprets the LARA aspects and communicates their intentions to the Weaving Engine. This engine will translate these intentions to specific actions that are applied on the internal program representation, the AST of Cetus. Cetus makes use of the developed transformation and analysis passes to change the program and send information back to the Weaving Engine.

% % % % % % % % % % % % % % % % %
% MONITORING STRATEGIES

\chapter{Monitoring Strategies}\label{chap:strats}

We consider different strategies for monitoring the range values of variables. Table \ref{tab:strats} summarizes the strategies that are detailed in the following sections. The LARA code for these strategies can be found in Appendix \ref{app:LARA}.

\begin{table}[t]
	\caption{A summary of the strategies used in this work. The last two columns indicate on which applications they were used.}
	\label{tab:strats}
	\centering
{\small 	\begin{tabular}{lp{.75\textwidth}}
	\toprule
	Strategy	& Description	\\
	\midrule
	ASCV3		& Monitors function parameters, assignments and return values					\\
	ASCV3\_s	& Similar to ASCV3, but deals with assignments to struct fields					\\
	FREQ		& Monitors variables with a certain percentage of the total assignments			\\
	FANIN       & Monitors variables whose assignments use a certain number of  variables		\\
	COMBAND 	& Intersection of the variable sets from FREQ and FANIN      					\\
	COMBOR		& Union of the variable sets from FREQ and FANIN     							\\
	\bottomrule
	\end{tabular}}
\end{table}

\section{Fault Detection Using Range Values}\label{sec:process}
The process through which we can detect a fault can be divided into two subprocesses. First, during the training subprocess, we learn ranges for each of the relevant variables on the application. Then, during the execution subprocess, and when the value of a variable changes, we can check if the range of that variable is violated, raising an exception and acting accordingly if needed. These two subprocesses are described with more detailed in the following sections.

\subsection{Training}

During this training process, we capture the ranges of function parameters, return values and variable assignments that are considered relevant. We do this, for each variable, \textit{M} times, where \textit{M} is the number of training executions. During each of these executions, we try to exercise a different part of the application. We do this, as learning a range from a single execution would severely constrain our ability to detect faults and lead to a very low detection accuracy. Using several ranges for the same variable, all resulting from different executions, allows us to relax the learned range and have a better representation of the global use of a variable inside the application.

The final product of this training phase is a set of ranges for each variable, for each of the examples executed. For instance, if we were to monitor 5 variables and perform 100 training executions, we would have $ 5 \times 100 = 500 $ ranges.

\subsection{Execution}

The idea behind the this subprocess is simple, we simply check if the value assigned to a variable during execution violates the range that we learned for that variable. In order to calculate the final range of a variable, we simply choose from all the learned ranges some percentage and merge them. This percentage can represent a single test, in which case we will have a single random training execution giving the variable range, or it can represent 100\% of these ranges and we will have the final range as relaxed as possible.

\begin{figure}[h!]
	\centering
	\includegraphics[width=0.9\textwidth]{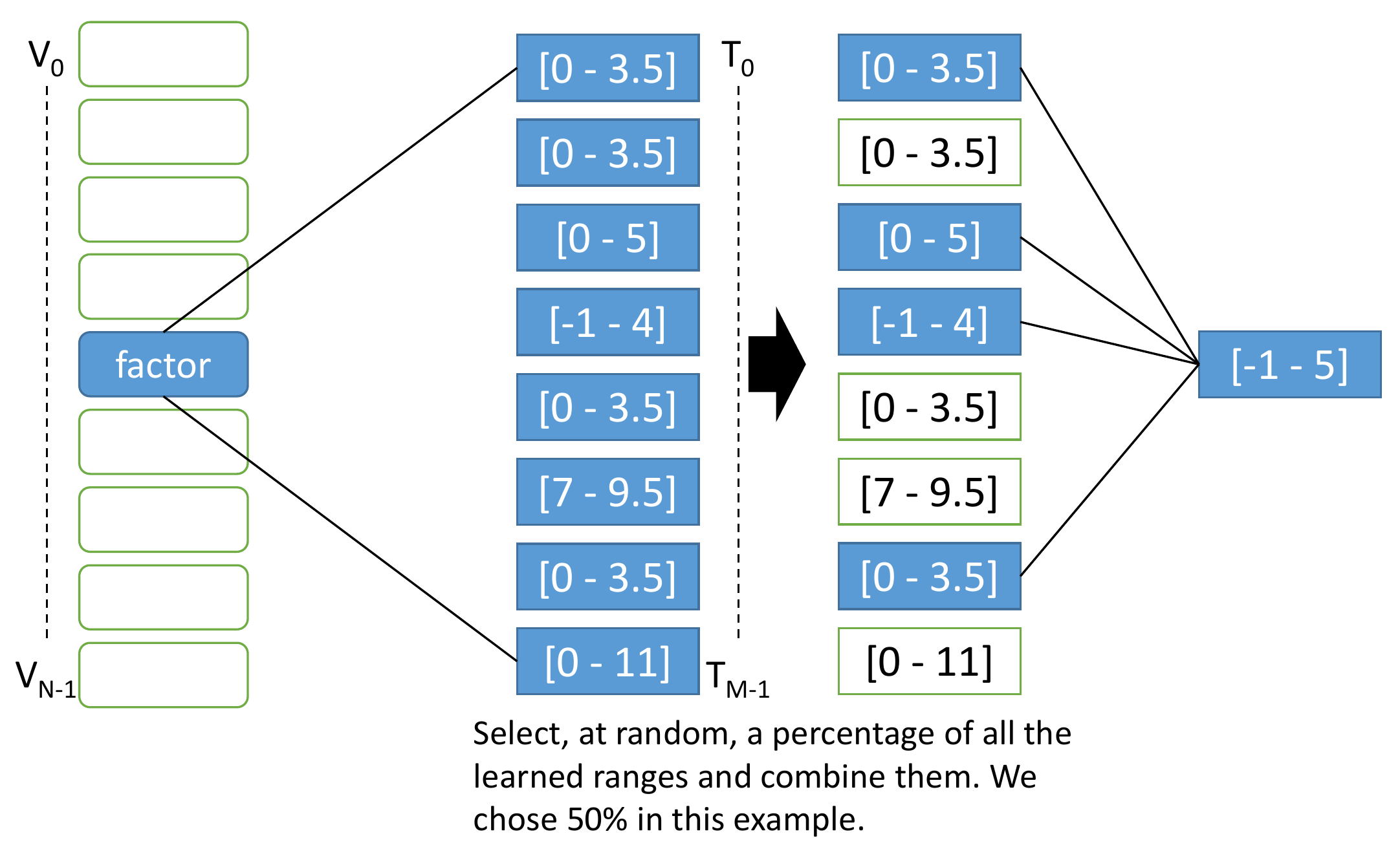}
	\caption{An example of the two subprocesses for a single variable. We perform \textit{M} training executions and choose 50\% of the resulting ranges to calculate the final range of the variable named \textit{factor}.}
	\label{fig:process}
\end{figure}

The two subprocesses are exemplified, for a single variable named \textit{factor} out of \textit{N} variables, on Figure \ref{fig:process}. In this example, we consider \textit{M} training executions, and choose 50\% of the ranges learned during those executions at random. These are used to calculate a relaxed range for the variable \textit{factor}.

\section{Main Strategy, Data Structures and Libraries}

\begin{figure}[h!]
	\centering
	\includegraphics[width=0.8\textwidth]{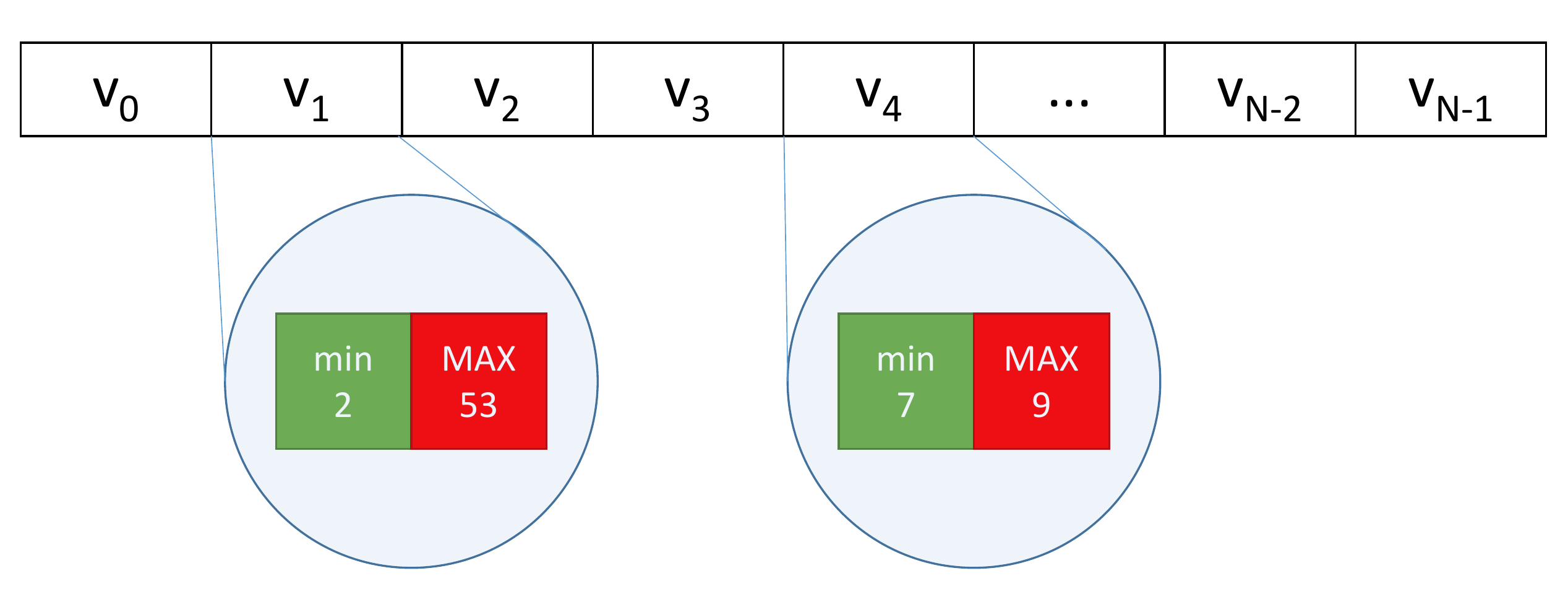}
	\caption{A representation of the array that stores the minimum and maximum values fo each monitored variable, considering \textit{N} variables.}
	\label{fig:array}
\end{figure}

The idea of the strategies described below is to use an array that can keep track of the range of every variable that we want to monitor. Each position on the array corresponds to a variable and the index of this position is calculated using only static information, during weaving time. This array contains pointers to structures that are able to save the range of a variable, as is depicted in Figure \ref{fig:array}. In practice, all we need is a minimum value and a maximum value and a way to update the range given a new value for a variable. This update is performed by the function \verb|update_range(Range* range, double value)|, part of a library developed to keep track of variable ranges during the instrumentation process. When we want to monitor a variable, we insert a call to the update function and give it the correct position on the array and the new value. The following example illustrates the an excerpt of the code that results from the instrumentation:

\begin{verbatim}
	a = b * c;
	update_range(&ranges[32], (double) a);
\end{verbatim}

The index of the array, 32 in this example, is unique to the variable \textit{a} on that particular function, and, as such, it can identify that variable uniquely. This is the index that is calculated at weaving time, using a JavaScript utility library provided with LARA. The value of the variable is the second argument to the \textit{range\_update} function and is casted to a double, the broadest scalar type available. The update function will simply update either the minimum or maximum value of the range struct with the value of \textit{a} (it can also update both minimum and maximum if this is the first update).

\section{ASCV3}
This strategy monitors function parameters, function return values and assignments to scalar variables. In order to monitor function parameters we insert, before the first statement of each function, a call to \textit{range\_update} per qualified parameter (we avoid pointers and arrays). This code is inserted right after the last declaration and it is the first code executed on the function. We also monitor each variable that is used on the return expression of a return statement. So, before each return statement of a given function, we insert a call to \textit{range\_update} for every variable that is used on the return expression, as such:

\begin{verbatim}
	range_update(..., (double) a);
	range_update(..., (double) b);
	return a + b;
\end{verbatim}

Finally, we monitor assignments to scalar variables, by calling \textit{range\_update} after each assignment. There are two exceptions. We do not monitor assignments performed on loop headers (e.g., the \verb|i=i+1| step expression) and we do not monitor assignments that are part of a conditional statement. By monitoring all these variables, we ensure that we take a good look at what is happening with each function, as we monitor its inputs and outputs, as well as most computations that happen inside it.

\section{ASCV3\_s}

This strategy performs the same instrumentation on the code as the ASCV3. However, because the previous strategy can not deal with assignments to struct member, it executes an additional preprocessing step. We apply a source code transformation that replaces assignments to struct members with equivalent code that is easier to instrument. The code:

\begin{verbatim}
	struct->member = value;
\end{verbatim}

is replaced with:

\begin{verbatim}
	temp_struct_member = value;
	struct->member = temp_struct_member;
\end{verbatim}

which can be easily instrumented as follows:

\begin{verbatim}
	temp_struct_member = value;
	range_update(..., temp_struct_member);
	struct->member = temp_struct_member;
\end{verbatim}

This makes it possible to instrument the usage of structs, which accounts for a large part of certain applications, using the previously defined strategy.

\section{FREQ}

Similar to the ASCV3\_s strategy, but now, we only monitor variables whose assignments account for a certain percentage of the total assignments. This strategy relies on a previous strategy that instruments the code in order to create a frequency report. The report contains the total number of assignments, as well as the number of assignments to each variable. With this information we can calculate each variable's share of the total assignments. This report is loaded and analyzed by the main strategy, ASCV3\_s, which chooses to monitor variables  whose share is higher than a given threshold. In this work we used the value 1\% as the threshold value. In the following example only the variables \textit{a} and \textit{b} would be chosen after their share was calculated:

\begin{verbatim}
a : 2.31%
b : 1.27%
c : 0.96%
d : 0.80%
...
g : 0.07%
\end{verbatim}

\section{FANIN}

Like the FREQ strategy, FANIN also makes use of an auxiliary strategy to generate a report that is used to choose only a subset of variables to monitor. This report contains, for each variable in each function, the maximum number of variables used on all of its assignments. The following code explains the fanin concept:

\begin{verbatim}
// fanin(a) = 3
a = b + c + d;

// fanin(a) = 2, but the maximum, 3, is chosen
a = b + c;

// fanin(e) = 1		
e = f + 0.1;

// fanin(g) = 1	
g = func(h) + i;
\end{verbatim}

The main strategy then selects variables that have a fanin higher than the input threshold, which, in this case, was 2. This means that, in our example, variables \textit{e} and \textit{g} would not be monitored.

\section{COMBAND and COMBOR}

COMBAND combines the previous strategies, performing an intersection of the resulting sets. The resulting set contains the variables that will be instrumented.

Much like the previous strategy, COMBOR also combines FREQ and FANIN but performs a union of the sets to produce the final set, the one that will be instrumented.

\section{Injected Faults}
\label{sec:faults}

The faults that were injected in the test versions of our example aaplications are based on the concept of mutants and they try to mimic faults introduced by humans writing code, e.g., using the operator $=$ instead of $==$. At any moment, only one fault is active.

\subsection{GZIP}

\paragraph{v1} The addition assignment operator ($+=$) is replaced by a simple assignment operator ($=$).

{\small \begin{verbatim}
#ifdef FAULTY_F_KL_6
    header_bytes = 2*sizeof(long);
#else
    header_bytes += 2*sizeof(long);
#endif
\end{verbatim}}

\paragraph{v2} This fault changes the operator $\leq$ for the operator $\geq$, effectivelly changing the decompression method.

{\small \begin{verbatim}
#ifdef FAULTY_F_KL_1
    if (compr_level >= 3) return deflate_fast();
#else
    if (compr_level <= 3) return deflate_fast(); /* for speed */
#endif
\end{verbatim}}

\paragraph{v3} If this fault is injected, the conditional expression is changed by removing the bitmask and the comparison with 0.

{\small \begin{verbatim}
#ifdef FAULTY_F_KL_5
    if (start[17])
#else
    if ((start[17] & 0xffff) != 0)
#endif
\end{verbatim}}

\paragraph{v4} When this fault is active the relational operator $\geq$ is replaced with $\leq$.

{\small \begin{verbatim}
#ifdef FAULTY_F_KL_1
    if (code <= 256) error("corrupt input.");
#else
    if (code >= 256) error("corrupt input.");
#endif
\end{verbatim}}

\paragraph{v5} This fault changes the values a variable that holds an error code.

{\small \begin{verbatim}
#define OK      0
#define ERROR   1
#define WARNING 2

/* ... */

#ifdef FAULTY_F_KL_2
    int err = -1;
#else
    int err = OK;
#endif
\end{verbatim}

\subsection{ABS}}

\paragraph{v1} This fault injects a logical negation operator ($!$), negating the conditional expression.

{\small \begin{verbatim}
#ifdef FAULT_1
    if (!(fac[j] < FACMIN))
#else
    if (fac[j] < FACMIN)
#endif
\end{verbatim}}

\paragraph{v2} When this fault is injected, we repalced the constant assignment.

{\small \begin{verbatim}
#ifdef FAULT_2
    int_T nx = 1;
#else
    int_T nx = 5;
#endif
\end{verbatim}}

\paragraph{v3} The arithmetic operator $/$ is replaced with $-$.

{\small \begin{verbatim}
#ifdef FAULT_3
    ztmp[i] = Delta[i]-hN;
#else
    ztmp[i] = Delta[i]/hN;
#endif
\end{verbatim}}

\paragraph{v4} The comparison operator $<$ is replaced with $>$.

{\small \begin{verbatim}
#ifdef FAULT_4
    localB->RelationalOperator1 = (*rtu_Input > 0.0);
#else
    localB->RelationalOperator1 = (*rtu_Input < 0.0);
#endif
\end{verbatim}}

\paragraph{v5} When this fault is active the assignment statement is removed.

{\small \begin{verbatim}
#ifdef FAULT_5
#else
    localXdot->WheelSpeed_CSTATE = 0.0;
#endif
\end{verbatim}

\section{Summary}
This chapter describes how we use range values to detect faults in programs written in C. There are two phases to our approach. Initially, we perform training runs where we learn the minimum and maximum value taken by each monitored variable on each of the runs. Then, we merge the ranges of a percentage of the runs to create the final range values for the monitored variables. We are then able to check if an assignment of a value violates the previously learned range.

\chapter{Experimental Results}\label{chap:exp}

In this chapter we demonstrate how we evaluated our approach. We used two applications and compared the results obtained with our approach with those obtained through the traditional approach of comparing the outputs of the applications. The first application is one of the first versions of GZIP, used to compress and decompress files and commonly packed on several Linux distributions. The second application is a Anti-lock Breaking System (ABS), which, given the initial velocities of the car and the wheels, calculates the distance needed to stop the car.

\section{Testing Process}

The test process compares the fault matrices that are generated using a perfect oracle, i.e. an unchanged, fault-free version of the application, with the fault matrices that are generated by our approach. A fault matrix contains, for each test execution and version of program, a boolean value which indicates whether the test is valid. An example of such a matrix can be seen in Figure \ref{fig:fault_matrix}.

\begin{figure}[h!]
	\centering
	\begin{math}
		\begin{matrix}
							& \mathbf{v_{1}}	& \mathbf{v_{2}}	& \mathbf{\cdots}	& \mathbf{v_{N}}	\\
			\mathbf{t_{1}}	& 1 				& 1					& \cdots			& 0					\\
			\mathbf{t_{2}}	& 0 				& 1					& \cdots			& 1					\\
			\mathbf{\vdots}	& \vdots			& \vdots			& \ddots			& \vdots			\\
			\mathbf{t_{M}}	& 0 				& 0					& \cdots			& 1					\\
		 \end{matrix}
	\end{math}
	\caption{An example of a fault matrix, where a value of 1 represents a passed test. In this example there are \textit{N} versions and \textit{M} test executions.}
	\label{fig:fault_matrix}
\end{figure}

To generate the perfect oracle matrix, we take the original version of the application and run all the tests. Then, we do the same for every version of the application with injected faults. In our case, there were 5 versions with faults for both example applications. By comparing the outputs of each version with the outputs of the original, we are able to build the fault matrix. If the outputs are the same then a value 1 is assigned and a value of 0 if the outputs are different. For instance, using the example in Figure \ref{fig:fault_matrix}, the output of the original version for the first test is equal to the output of versions 1 and 2 for the same test, while it is different from the output of version \textit{N} (elements \{1,1\}, \{1,2\} and \{1,N\}). 

In order to evaluate our approach, we also generate fault matrices but refrain from using the fault-free version. Instead, we use the training and execution processes described in Section \ref{sec:process}. We learn, for each relevant variable according to our strategy, a range and verify if that range is violated. On a deployed application this test could be performed online, as it is a simple assertion. In our case, for each of the test executions, we generated a new range, which should be contained within the previously learned range. The learned range can use any percentage of the ranges collected during the training phase (we used 5\%, 10\%, 20\%, 30\%, 40\%, 50\%, 60\%, 70\%, 80\%, 90\% and 100\%). For each element of the matrix, i.e., for a test and a version, if the test range of \textbf{any} of the monitored variables is not contained on the learned range, the test is considered failed and a value of 0 is assigned. Otherwise, we assume the test is valid and assign a value of 1.

The evaluation presented in this chapter compares, for both example applications, the fault matrices generated from the perfect oracle with the matrices generated using our approach for several learning percentages. 

\section{Results}
The tables in this section represent the accuracy of each of the tested strategies. This accuracy represents how similar the fault matrices generated by our testing process are to the fault matrix generated by the usage of the perfect oracle. For instance if a fault matrix generated by our approach is completely equal to the matrix generated from the original application we would have an accuracy of 100\%. Conversely, an accuracy of 0\% would mean that the matrices are the complete opposite of each other, as they only contain boolean values.

The tables presented in this section illustrate, for each of the tested strategies (described in Chapter \ref{chap:strats}), the best accuracy achieved and the corresponding learning percentage. If two different learning percentages achieve the same accuracy, e.g. both 90\% and 100\% manage to achieve 65\% accuracy, then we show the smallest learning percentage as the best.

\subsection{ABS Results}

The tests performed on the ABS application show an overall low accuracy, as can be observed in Table \ref{tab:abs_overall_results}. There is an upper bound around 54\% accuracy which will be explained later. There isn't a difference between the top accuracies achieved by the different strategies, although we can consider FREQ as the best strategy.

FREQ and COMBAND, which share some of the monitored variables are able to reach their best detection accuracy using a smaller training percentage than the rest of the strategies. This means that these strategies are able to find a set of key variables to monitor.

These results remain difficult to explain, since the FREQ and FANIN share only a small number of variables and their best accuracy is identical. We believe that this may be related to the existence of a large number of false positive results, which is presented ahead.

%%%%%%%%%%%%%%%%%%%%%%%%%%%%%%%%%%%%%%%%%%%%%%%%%%%%%
%%%%   overall results for the ABS application   %%%%
%%%%%%%%%%%%%%%%%%%%%%%%%%%%%%%%%%%%%%%%%%%%%%%%%%%%%
\begin{table}[h]
	\caption{Overall results for the ABS application.}
	\label{tab:abs_overall_results}
	\centering
{\small 	\begin{tabular}{lrr}
	\toprule
	Strategy	& Best Accuracy (\%)	& Best Percentage	\\
	\midrule
	ASCV3		& 54.40					& 80	\\
	ASCV3\_s	& 54.40					& 100	\\
	FREQ		& 54.93    				& 10	\\
	FANIN       & 54.40    				& 100	\\
	COMBAND 	& 54.40    				& 10	\\
	COMBOR		& 54.40    				& 100	\\
	\bottomrule
	\end{tabular}}
\end{table}

%%%%%%%%%%%%%%%%%%%%%%%%%%%%%%%%%%%%%%%%%%%%%%%%%%
%%%%   c, fp and fn for the ABS application   %%%%
%%%%%%%%%%%%%%%%%%%%%%%%%%%%%%%%%%%%%%%%%%%%%%%%%%
We can see, in Figure \ref{fig:abs_cfpcn}, the evolution of the number of correct predictions, false positives and false negatives for each strategy and for each training percentage used on the ABS application. It is evident that there is a large number of false positive results throughout the entire range of strategies and percentages. The number of false negatives decreases as we increase the training percentage, which is a trend that can be seen on all strategies. The percentage of correct results seems to increase slightly, while the number of false positives remains almost unchanged. It is worth noting that the strategies FREQ and COMBAND only present false negative results for the 5\% training percentage, which are eliminated when we train with at least 10\% of the data.

%%%%%%%%%%%%%%%%%%%%%%%%%%%%%%%%%%%%%%%%%%%%%%%%%%
%%%%   c, fp and fn for the ABS application   %%%%
%%%%%%%%%%%%%%%%%%%%%%%%%%%%%%%%%%%%%%%%%%%%%%%%%%
\begin{figure}
        \centering
		\includegraphics{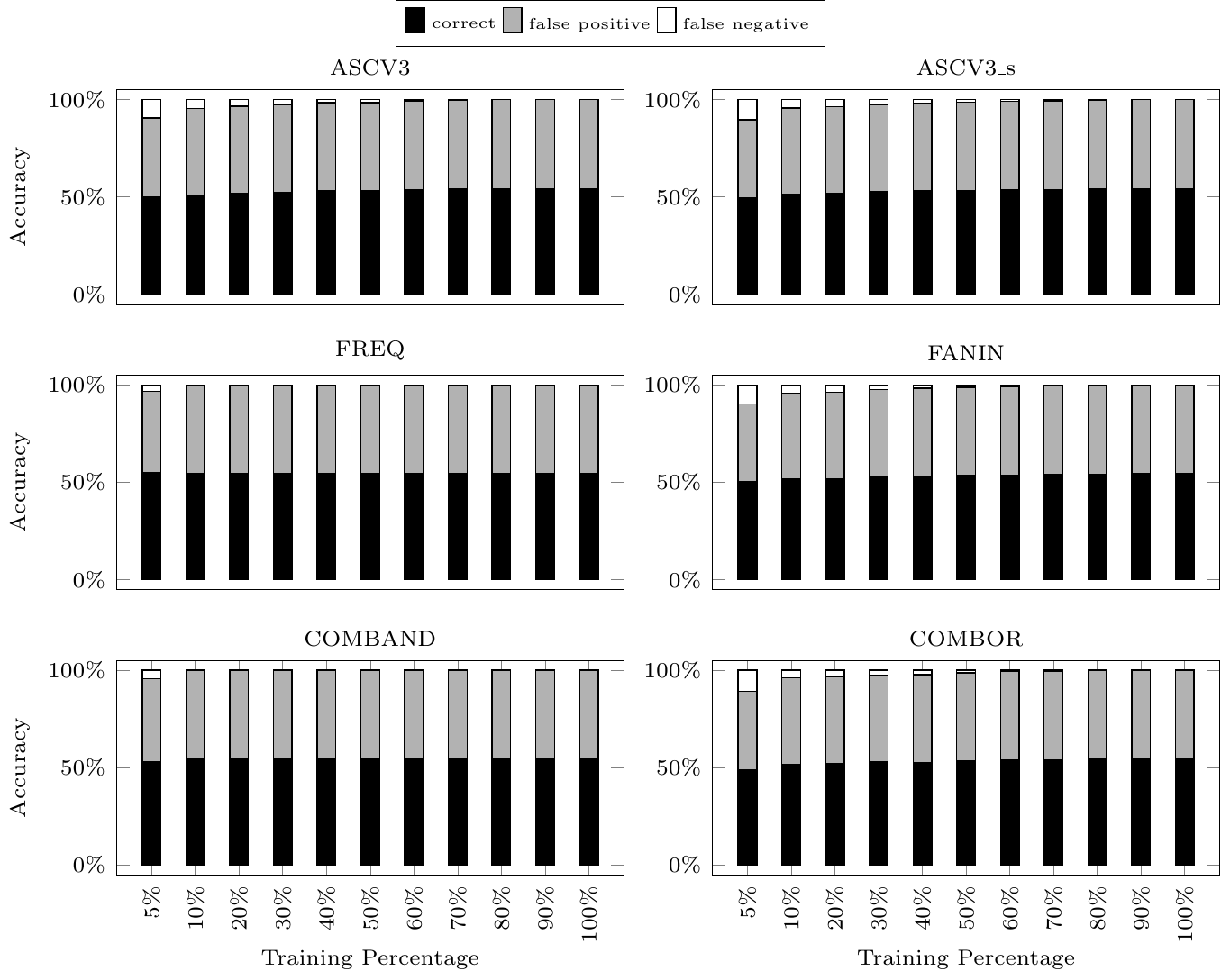}
        \caption{The percentage of correct predictions, false positives and false negatives for all the tested training percentages for the ABS application.}
        \label{fig:abs_cfpcn}
\end{figure}

The test results per version for the ABS application are shown in Table \ref{tab:abs_version_results}. Versions two and five have poor results, achieving a top accuracy that ranges from 2.4\% to a maximum of 12.27\%. These results were obtained with only 5\% usage of the training data.

The fault injected in the second version changes the initialization of a variable that is immediately used to perform two \textit{memcpy} function calls, as the number of bytes to copy. Because the fault reduces the number from 5 to 1, the number of copied bytes is smaller and we are left with two arrays with uninitialized positions. Furthermore, this variable is also used as the upper bound of a \textit{FOR} loop that initializes a third array through pointer increments. Because our current strategies do not monitor the values inside arrays, we fail to capture this change with our approach. We tried, however to monitor the control variable of this loops, by using a strategy that monitors loop control variables before and after the execution of loops. Any value that this variable takes inside the loop body may not be captured. The results were not better then those already obtained with the other strategies. We achieved a top accuracy of 54.4\% when using 90\% of the training data.
	
The poor accuracy of the fifth version may also be explained by the type of fault that was injected. This fault removes an assignment to a variable that is used to calculate an internal derivative for a representation of the speed of a wheel. This directly affects one of the outputs of the program, meaning that it is easy to detect when simply comparing the outputs of the original and mutant versions. However, this fault changes only a very reduced number of variables inside the program (which may not even be monitored, if they are arrays or pointers), hampering our ability to detect it with our range approach.

We achieve good accuracy on the first version for all the strategies. Although this is good, we expected a larger difference from the first strategy to the second, since the later monitors more than the double of variables than the former (which can be seen in Table \ref{tab:abs_jps}). Finally, we achieve perfect results for versions three and four. This is achieved with a large percentage of training data, except for the FREQ and FANIN strategies, which need only 10\%.

%%%%%%%%%%%%%%%%%%%%%%%%%%%%%%%%%%%%%%%%%%%%%%%%%%%%%%%%%
%%%%   per version results for the ABS application   %%%%
%%%%%%%%%%%%%%%%%%%%%%%%%%%%%%%%%%%%%%%%%%%%%%%%%%%%%%%%%
\begin{table}[h]
	\caption{Per version results for the ABS application.}
	\label{tab:abs_version_results}
	\centering
{\scriptsize \begin{tabular}{lrrrrrcrrrrr}
\toprule
	& \multicolumn{5}{c}{Best Accuracy (\%)}	& & \multicolumn{5}{c}{Best Percentage} \\
\cmidrule(r){2-6} \cmidrule(r){8-12}
Strategy		& v1	& v2	& v3	& v4	& v5	&& v1	& v2	& v3	& v4	& v5	\\
\midrule
	ASCV3		& 74.80	& 5.33	& 100	& 100	& 10.40	&& 10	& 5		& 100	& 90	& 5	\\
	ASCV3\_s	& 75.33	& 7.20	& 100	& 100	& 10.53	&& 10	& 5		& 90	& 100	& 5	\\
	FREQ		& 72.00	& 4.53	& 100	& 100	& 12.27	&& 10	& 5		& 10	& 10	& 5	\\
	FANIN		& 75.73	& 6.67	& 100	& 100	& 11.73	&& 10	& 5		& 90	& 100	& 5	\\
	COMBAND		& 72.00	& 2.4	& 100	& 100	& 2.7	&& 10	& 5		& 10	& 10	& 5	\\
	COMBOR		& 75.60	& 7.87	& 100	& 100	& 9.33	&& 10	& 5		& 100	& 100	& 5	\\
\bottomrule
\end{tabular}}
\end{table}

%%%%%%%%%%%%%%%%%%%%%%%%%%%%%%%%%%%%%%%%%%%%%%%%%%%%%%%%%
%%%%   changing the constant value on the v2 fault   %%%%
%%%%%%%%%%%%%%%%%%%%%%%%%%%%%%%%%%%%%%%%%%%%%%%%%%%%%%%%%
We performed one final test on this application. We changed the faulty value of the constant that is redefined when the fault is active. So, instead of testing just the value 1, we tested some others. The accuracy results are presented in Table \ref{tab:abs_v2_values}. The only strategies presented are the ones where the change had a noticeable impact.

\begin{table}[h]
	\caption{Accuracy results obtained when different faulty constant values are injected in the second version. The original faulty value is highlighted.}
	\label{tab:abs_v2_values}
	\centering
{\scriptsize \begin{tabular}{lrrrrr}
\toprule
	& \multicolumn{5}{c}{Faulty Values}	 \\
\cmidrule(r){2-6}
Strategy		& \textbf{1}		& 0		& -1	& 10	& 20		\\
\midrule
	ASCV3		& \textbf{5.33}		& 19.6	& 19.47	& 17.07	& 22.27		\\
	ASCV3\_s	& \textbf{7.20}		& 23.87	& 24.93	& 21.87	& 22.27		\\
\bottomrule
\end{tabular}}
\end{table}

We can see that out of all the tested strategies, only 2 of them produced different results when the constant value is changed. It is worth noting that FREQ, the best overall strategy for this application and COMBOR, the best strategy for this version, failed to detect this change. This is likely due to the fact that the first two strategies monitor a greater number of variables, including those whose range changes with this fault.

In the function where this fault is located, only a single variable is monitored and it is used to hold a time instant. This variable however is not directly changed using the faulty variable. It changes inside a loop that performs differently based on the faulty variable. This fault propagates and ends up changing the value of the only function parameter, a pointer to a struct whose fields are updated inside the function. Although the fault is located here, it is not detected with our ranges in this function. It propagates and is detected somewhere else.

Boundary values, such as the tested 0 and -1, are likely to have a large impact on the control flow of the program. The variable where the fault is injected is mainly used to control loops that initialize and update arrays. This means that it is possible to completely skip the execution of certain blocks of the application. The fact that larger values also produce more accurate results may be explained by the type of application we are using. This is an program that relies heavily on numerical calculation and certain control flow structures can be heavily affected by numerical values, for instance, when stopping a calculation inside a loop if a certain value is above a predefined threshold. Because of the nature of such programs, changing a variable like we do with that fault, can have an exponential propagation on ranges and also control flow structures.

\subsection{GZIP Results}

Table \ref{tab:gzip_overall_results} presents the overall the prediction accuracy for the GZIP application. This accuracy is good, with every strategy correctly predicting around the 80\% of the test results. The strategies FREQ and COMBAND achieve the best accuracy and this last only needs to use 20\% of the training data to achieve it.

There doesn't seem to be a very large difference between the ability of different strategies to detect faults. Also, there isn't much of a difference between the first two strategies, as including the assignments to struct fields has an almost null effect on the number of monitored variables for the GZIP application (as is shown in Table \ref{tab:gzip_jps}).

FANIN and COMBOR, although managing to achieve better performance than the first two strategies, are worse than FREQ and COMBAND. It seems that FREQ finds a better set of variables to monitor for this particular application. Moreover, the same happens with COMBAND, as it does not use any variables that are not already included in FREQ.

%%%%%%%%%%%%%%%%%%%%%%%%%%%%%%%%%%%%%%%%%%%%%%%%%%%%%%
%%%%   overall results for the GZIP application   %%%%
%%%%%%%%%%%%%%%%%%%%%%%%%%%%%%%%%%%%%%%%%%%%%%%%%%%%%%
\begin{table}[h]
	\caption{Overall results for the GZIP application.}
	\label{tab:gzip_overall_results}
	\centering
{\small 	\begin{tabular}{lrr}
	\toprule
	Strategy	& Best Accuracy (\%)	& Best Percentage	\\
	\midrule
	ASCV3		& 80.37					& 100	\\
	ASCV3\_s	& 80.37					& 100	\\	
	FREQ		& 83.18					& 100	\\
	FANIN       & 82.43					& 90	\\
	COMBAND 	& 83.18					& 20	\\
	COMBOR		& 82.43					& 100	\\
	\bottomrule
	\end{tabular}}
\end{table}

%%%%%%%%%%%%%%%%%%%%%%%%%%%%%%%%%%%%%%%%%%%%%%%%%%%
%%%%   c, fp and fn for the GZIP application   %%%%
%%%%%%%%%%%%%%%%%%%%%%%%%%%%%%%%%%%%%%%%%%%%%%%%%%%
Figure \ref{fig:gzip_cfpcn} illustrates the evolution of the percentage of correct and incorrect predictions over the used training percentages for each of the tested strategies on the GZIP application. The charts start at 60\% to give a clearer indication of each of the percentages.

The ASCV3 and ASCV3\_s follow similar patterns, starting with almost the same number of false positives and false negatives. Then, as the percentage of used training data increases, the percentage of correct predictions increases, the false negative becomes almost null and the false positive rate increases slightly.

Although, at a training percentage of 100\%, the FANIN strategy manages to achieve roughly the same prediction accuracy as the FREQ strategy , the later is able to achieve for a smaller amount of training data. This is an indication that the variables chosen by this strategy are better, as they at least require a smaller amount of training data, improving the quickness of our testing process. The same happens with the last two strategies, COMBAND and COMBOR: COMBAND does not add variables other than those used by FREQ, achieving the same early results and COMBOR adds the variables from FANIN, which seem to dilute the effect of the variables selected by FREQ.

%%%%%%%%%%%%%%%%%%%%%%%%%%%%%%%%%%%%%%%%%%%%%%%%%%%
%%%%   c, fp and fn for the GZIP application   %%%%
%%%%%%%%%%%%%%%%%%%%%%%%%%%%%%%%%%%%%%%%%%%%%%%%%%%
\begin{figure}
        \centering
        \includegraphics{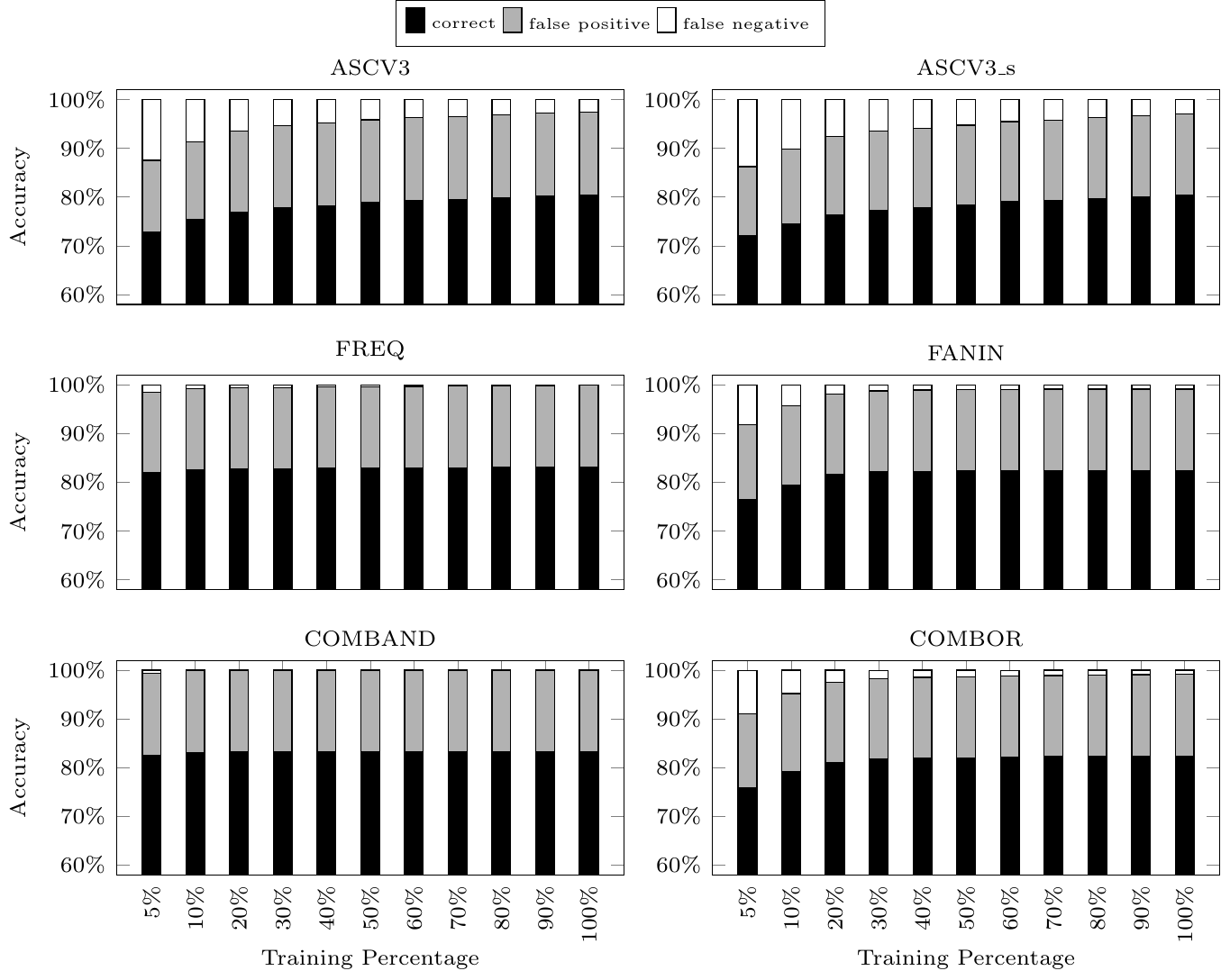}
        \caption{The percentage of correct predictions, false positives and false negatives for all the tested training percentages for the GZIP application.}
        \label{fig:gzip_cfpcn}
\end{figure}

When looking at the different versions, shown in Table \ref{tab:gzip_version_results}, the results are overall great with the exception of version number two, where the highest accuracy achieved is 24.17\% and all the strategies achieve an accuracy between 21\% and 24\%.

The bad results on the second version of GZIP can be explained by the type of fault that was injected (please see Section \ref{sec:faults}). This fault changes the compression method used to a faster one, that produces a different output. The value change in this control variable will not impact a lot of other variables inside the application code but will have a direct impact on the output, much like the case with the fifth version of the ABS application. This means that when testing for the presence of a fault by comparing the output we will easily detect this, but it becomes harder when using our range strategies.

For every version but the second one, the last four strategies are better than the first two, which monitor a lot more variables. FREQ and COMBAND achieve the best results for every version except the second. Again, FREQ selects a good set of variables, which is further limited when applying the set intersection operation with the set from FANIN to generate the set from the COMBAND strategy. COMBAND also separates itself from the other strategies because it manages to achieve roughly the same accuracy but using a significantly smaller amount of training data. Once more, COMBOR seems to have the effect of diluting the variables chosen by FREQ, although the difference in the prediction accuracy is small.

Similarly to what we did with the ABS application we tested a new strategy that augments the best strategy and adds support to monitor loop control variables. Once more, the results obtained with this strategy were on par with those already obtained. The top accuracy was 80.37\% with 100\% of the training data.

%%%%%%%%%%%%%%%%%%%%%%%%%%%%%%%%%%%%%%%%%%%%%%%%%%%%%%%%%%
%%%%   per version results for the GZIP application   %%%%
%%%%%%%%%%%%%%%%%%%%%%%%%%%%%%%%%%%%%%%%%%%%%%%%%%%%%%%%%%
\begin{table}[h]
	\caption{Per version results for the GZIP application.}
	\label{tab:gzip_version_results}
	\centering
{\scriptsize \begin{tabular}{lrrrrrcrrrrr}
\toprule
	& \multicolumn{5}{c}{Best Accuracy (\%)}	& & \multicolumn{5}{c}{Best Percentage} \\
\cmidrule(r){2-6} \cmidrule(r){8-12}
Strategy		& v1	& v2	& v3	& v4	& v5	&& v1	& v2	& v3	& v4	& v5	\\
\midrule
	ASCV3		& 92.99	& 23.27	& 96.73	& 97.20	& 96.26	&& 100	& 5		& 100	& 100	& 100	\\
	ASCV3\_s	& 92.99	& 24.17	& 96.73	& 97.20	& 96.26	&& 100	& 5		& 100	& 100	& 100	\\
	FREQ		& 95.86	& 22.12	& 100	& 99.53	& 99.07	&& 100	& 5		& 100	& 100	& 100	\\
	FANIN		& 95.33	& 21.42	& 99.07	& 99.53	& 98.13	&& 50	& 5		& 50	& 90	& 100	\\
	COMBAND		& 96.26	& 21.14	& 100	& 99.53	& 99.07	&& 5	& 5		& 20	& 20	& 20	\\
	COMBOR		& 95.33	& 22.73	& 99.07	& 99.53	& 98.13	&& 100	& 5		& 100	& 100	& 100	\\
\bottomrule
\end{tabular}}
\end{table}

\subsection{Instrumentation Results}

% join points
Tables \ref{tab:abs_jps} and \ref{tab:gzip_jps} illustrate the number of join points that are instrumented for the ABS and GZIP applications respectively. A join point is a point in the code that we may want to instrument. The tables show the number of selected points, i.e., the those considered for instrumentation) and the number of advised points (i.e., those that were actually instrumented).

One thing that is immediately clear is that this instrumentation would be a rather difficult task to perform manually. Although, in some cases, we instrument a small set of variables, we always analyze a large number. Going manually through each of these, would be not only be time-consuming, but also error-prone. This is the case for every code transformation that is applied manually to more than a few lines of code.

In the ABS application, the last version has less selected points because those are removed by the injected fault (described in Section \ref{sec:faults}) that removes a statement from the application code. There is a large difference in the number of selected and advised points between the first two strategies. The second strategy, ASCV3\_s, decomposes assignments to struct fields and this allows us to instrument these assignments as if they were scalar variables. All the strategies that follow use this same transformation, which results in a larger amount of monitored points, as the ABS application makes heavy use of structs for its execution. On the other hand, we can see that, with GZIP, we only monitor one more variable if we consider the usage of structs in the code. Hence, the last four strategies, just like the first one, do not consider structs in the tests for this application.

%%%%%%%%%%%%%%%%%%%%%%%%%%%%%%%%%%%%%%%%%%%%%%%%%%%%%
%%%%   join point data for the ABS application   %%%%
%%%%%%%%%%%%%%%%%%%%%%%%%%%%%%%%%%%%%%%%%%%%%%%%%%%%%
\begin{table}[h]
	\caption{Number of advised and selected join points per version of the ABS application. The columns labeled with \textit{A} indicate advised join points while the columns labeled with \textit{S} indicate selected join points.}
	\label{tab:abs_jps}
	\centering
{\scriptsize \begin{tabular}{lrrcrrcrrcrrcrr}
\toprule
	& \multicolumn{2}{c}{v1} & & \multicolumn{2}{c}{v2}  & & \multicolumn{2}{c}{v3} & & \multicolumn{2}{c}{v4} & & \multicolumn{2}{c}{v5}\\
\cmidrule(r){2-3} \cmidrule(r){5-6} \cmidrule(r){8-9} \cmidrule(r){11-12} \cmidrule(r){14-15}
Strategy	& A & S && A & S && A & S && A & S && A & S \\
\midrule
ASCV3		& 52 	& 1340 && 52	& 1340 && 52	& 1340 && 52	& 1340 && 52	& 1339 \\
ASCV3\_s	& 129	& 1742 && 129	& 1742 && 129	& 1742 && 129	& 1742 && 129	& 1738 \\
FREQ		& 33 	& 1742 && 33	& 1742 && 33	& 1742 && 33	& 1742 && 33	& 1738 \\
FANIN		& 21 	& 1742 && 21	& 1742 && 21	& 1742 && 21	& 1742 && 21	& 1738 \\
COMBAND		& 9  	& 1742 && 9 	& 1742 && 9		& 1742 && 9		& 1742 && 9		& 1738 \\
COMBOR		& 45 	& 1742 && 45	& 1742 && 45	& 1742 && 45	& 1742 && 45	& 1738 \\
\bottomrule
\end{tabular}}
\end{table}

%%%%%%%%%%%%%%%%%%%%%%%%%%%%%%%%%%%%%%%%%%%%%%%%%%%%%%
%%%%   join point data for the GZIP application   %%%%
%%%%%%%%%%%%%%%%%%%%%%%%%%%%%%%%%%%%%%%%%%%%%%%%%%%%%%
\begin{table}[h]
	\caption{Number of advised and selected join points per version of the GZIP application. The columns labeled with \textit{A} indicate advised join points while the columns labeled with \textit{S} indicate selected join points.}
	\label{tab:gzip_jps}
	\centering
{\scriptsize \begin{tabular}{lrrcrrcrrcrrcrr}
\toprule
	& \multicolumn{2}{c}{v1} & & \multicolumn{2}{c}{v2}  & & \multicolumn{2}{c}{v3} & & \multicolumn{2}{c}{v4} & & \multicolumn{2}{c}{v5}\\
\cmidrule(r){2-3} \cmidrule(r){5-6} \cmidrule(r){8-9} \cmidrule(r){11-12} \cmidrule(r){14-15}
Strategy	& A & S && A & S && A & S && A & S && A & S \\
\midrule
ASCV3		& 395	& 4627 && 482	& 5379		&& 458	& 5009 && 461	& 5053 && 450	& 5212 \\
ASCV3\_s	& 396	& 4714 && 483	& 5466		&& 459	& 5096 && 462	& 5140 && 451	& 5299 \\
FREQ		& 24	& 4627 && 22	& 5379		&& 22	& 5009 && 22	& 5053 && 22	& 5212 \\
FANIN		& 48	& 4627 && 54	& 5379		&& 50	& 5009 && 50	& 5053 && 51	& 5212 \\
COMBAND		& 5 	& 4627 && 5		& 5379		&& 5	& 5009 && 5		& 5053 && 5		& 5212 \\
COMBOR		& 67	& 4627 && 71	& 5379		&& 67	& 5009 && 67	& 5053 && 68	& 5212 \\
\bottomrule
\end{tabular}}
\end{table}

% instrumentation times
It is possible to see, in Table \ref{tab:abs_times} and Table \ref{tab:gzip_times}, the instrumentation times, in milliseconds, for the ABS and GZIP versions. The ABS table contains only the single version where the five different faults were injected. For this application, the instrumentation times vary between around 3 seconds and 4.8 seconds. The instrumentation times are somewhat close, which might mean that most of this time is related to the initialization of the Java Virtual Machine, as the weaving compiler, MANET, is a Java application. The FREQ, FANIN, COMBAND and COMBOR strategies take longer as they need to load reports and choose the variables that will be instrumented based on their contents. Although the instrumentation times are longer, the same pattern can be seen for the GZIP application, in Table \ref{tab:gzip_times}. The longer times are explained because this is a much larger application. The fastest weaving took around 4 seconds, while the longest took around 7 seconds.

%%%%%%%%%%%%%%%%%%%%%%%%%%%%%%%%%%%%%%%%%%%%%%%%%%%%%%%%%%%
%%%%   instrumentation times for the ABS application   %%%%
%%%%%%%%%%%%%%%%%%%%%%%%%%%%%%%%%%%%%%%%%%%%%%%%%%%%%%%%%%%
\begin{table}[h]
	\caption{Instrumentation times for the ABS application.}
	\label{tab:abs_times}
	\centering
{\small 	\begin{tabular}{lr}
	\toprule
	Strategy	& Time (ms)	\\
	\midrule
	ASCV3		& 3067		\\
	ASCV3\_s	& 3976		\\
	FREQ		& 4814		\\
	FANIN       & 4739		\\
	COMBAND 	& 4717		\\
	COMBOR		& 4774		\\
	\bottomrule
	\end{tabular}}
\end{table}

%%%%%%%%%%%%%%%%%%%%%%%%%%%%%%%%%%%%%%%%%%%%%%%%%%%%%%%%%%%%
%%%%   instrumentation times for the GZIP application   %%%%
%%%%%%%%%%%%%%%%%%%%%%%%%%%%%%%%%%%%%%%%%%%%%%%%%%%%%%%%%%%%
\begin{table}[h]
	\caption{Instrumentation times for the GZIP application.}
	\label{tab:gzip_times}
	\centering
{\small 	\begin{tabular}{lrrrrr}
	\toprule
					& \multicolumn{5}{c}{Time (ms)}	 \\
					\cmidrule(r){2-6}
	Strategy		& v1	& v2	& v3	& v4	& v5		\\
	\midrule
		ASCV3		& 3998	& 4694	& 4224	& 4286	& 4420		\\
		ASCV3\_s	& 4035	& 4012	& 4297	& 4578	& 4885		\\		
		FREQ		& 6356	& 6976	& 6659	& 6756	& 7065		\\
		FANIN		& 6257	& 7041	& 6624	& 6691	& 6992		\\
		COMBAND		& 6284	& 6979	& 6650	& 6790	& 7039		\\
		COMBOR		& 6292	& 7087	& 6777	& 6834	& 7035		\\
	\bottomrule
	\end{tabular}}
\end{table}

% slowdown
Table \ref{tab:abs_slowdown} and Table \ref{tab:gzip_slowdown} present the slowdowns of the instrumented versions of ABS and GZIP respectively. In the ABS application (for which we make no distinction between versions) the slowdowns range from 0.86 to 0.96, meaning that we get execution times close to the original, non-instrumented version. Even though, in some strategies, we are monitoring a large number of variables, the slowdown is still acceptable and the instrumented application runs in useful time. The same can be observed for the GZIP application. The slowdowns range from 0.77 (for the ASCV3\_s strategy on the first version) to 0.91 (for the COMBAND strategy on the second version). One of the important features of any instrumentation approach is the increase in the execution time of the instrumented application. Our approach can collect the data needed for the analysis without even doubling this execution time.

%%%%%%%%%%%%%%%%%%%%%%%%%%%%%%%%%%%%%%%%%%%%%%%%%%%%%%%%%%
%%%%   slowdown of the instrumented ABS application   %%%%
%%%%%%%%%%%%%%%%%%%%%%%%%%%%%%%%%%%%%%%%%%%%%%%%%%%%%%%%%%
\begin{table}[h]
	\caption{Slowdown of the instrumented ABS application when compared to the original.}
	\label{tab:abs_slowdown}
	\centering
{\small 	\begin{tabular}{lr}
	\toprule
	Strategy	& Slowdown	\\
	\midrule
	ASCV3		& 0.86		\\
	ASCV3\_s	& 0.86		\\
	FREQ		& 0.89		\\
	FANIN       & 0.91		\\
	COMBAND 	& 0.89		\\
	COMBOR		& 0.96		\\
	\bottomrule
	\end{tabular}}
\end{table}

%%%%%%%%%%%%%%%%%%%%%%%%%%%%%%%%%%%%%%%%%%%%%%%%%%%%%%%%%%%
%%%%   slowdown of the instrumented GZIP application   %%%%
%%%%%%%%%%%%%%%%%%%%%%%%%%%%%%%%%%%%%%%%%%%%%%%%%%%%%%%%%%%
\begin{table}[h]
	\caption{Slowdown for each version of the instrumented GZIP application when compared to the original.}
	\label{tab:gzip_slowdown}
	\centering
{\small 	\begin{tabular}{lrrrrr}
	\toprule
				& \multicolumn{5}{c}{Slowdown}	 \\
				\cmidrule(r){2-6}	
	Strategy	& v1	& v2	& v3	& v4	& v5	\\
	\midrule			
	ASCV3		& 0.86	& 0.83	& 0.84	& 0.84	& 0.86	\\
	ASCV3\_s	& 0.77	& 0.73	& 0.80	& 0.84	& 0.85	\\	
	FREQ		& 0.86	& 0.86	& 0.85	& 0.84	& 0.85	\\
	FANIN		& 0.89	& 0.85	& 0.82	& 0.89	& 0.85	\\
	COMBAND		& 0.89	& 0.91	& 0.89	& 0.90	& 0.87	\\
	COMBOR		& 0.87	& 0.87	& 0.86	& 0.86	& 0.85	\\
	\bottomrule
	\end{tabular}}
\end{table}

\subsection{False Positives Analysis}

% % % % % % % % % % % % % % % % % % % % % % % % % % % % % %
% metrics explanation
% % % % % % % % % % % % % % % % % % % % % % % % % % % % % %
In this section, several metrics are used. The next paragraphs explain their meaning and how they are calculated.

\begin{description}
	\item[tp:] true positive count
	\item[fp:] false positive count
	\item[tn:] true negative count
	\item[fn:] false negative count
	\item[ACC:] Accuracy. The metric previously used.
	\item[PPV:] Positive Predictive Value. The percentage of correct predictions where we say the test will pass. Also known as Precision. It is calculated as:
		\begin{math}
		PPV = \dfrac{tp}{tp + fp}
		\end{math}
	\item[NPV:] Negative Predictive Value. The percentage of correct predictions where we say the test will fail. It is calculated as:
		\begin{math}
		NPV = \dfrac{tn}{tn + fn}
		\end{math}
	\item[TPR:] True Positive Rate. The percentage of the positive instances that we manage to identify. Also known as Recall. It is calculated as:
		\begin{math}
		TPR = \dfrac{tp}{tp + fn}
		\end{math}
	\item[TNR:] True Negative Rate. The percentage of the negative instances that we manage to identify. It is calculated as:
		\begin{math}
		TNR = \dfrac{tn}{fp + tn}
		\end{math}
\end{description}

Whenever the expressions \textit{negative prediction} or \textit{positive predictions} are used, we mean a prediction, made by our approach, that the test will fail or pass respectively. They do not mean that the prediction is incorrect or correct. The same applies for the expressions \textit{negative instance} or \textit{positive instance}. They only mean that particular case is a failed or passed test in the original fault matrix.

% % % % % % % % % % % % % % % % % % % % % % % % % % % % % %
% ABS
% % % % % % % % % % % % % % % % % % % % % % % % % % % % % %

The fact that, for the v2 and v5 versions of ABS, the best results were obtained with the lowest training percentage (5\%), is due to the fact that anything above that has 0\% accuracy. As we increase the training percentage,  we move farther away from the original fault matrix.

Hereupon, we decided to explore the hypothesis that our approach does not fit the code of the ABS application. We assumed that the problem does not exist only on versions v2 and v5 (that show the worst accuracy), but that it presents itself on every tested version. This could be explained if our approach was simply passing every single one of the tests and producing a high number of false positive results. The problem with detecting these false negatives arises when the original fault matrix has a small number of failed tests, which is the case.

We decided to look at the original fault matrix from the ABS application. The cases with highest number of false positive results are mainly in the second and fifth versions, which, according to the original matrix, is where the largest number of failed tests occurs. If, as we hypothesized, our approach was blindly passing all tests, then our accuracy should be similar to the percentage of passed tests in the original fault matrix. We calculated this percentage for the ABS program and it has a value of 54\%. This result not only proves the hypothesis that our approach does not fit ABS, but it also explains the 54\% upper bound limit that was previously presented in the results.

Table \ref{tab:abs_1s} presents the best accuracy achieve for each version of the ABS as well as the percentage of passed tests in the original BAS fault matrix. Versions three and four have 100\% accuracy and all the tests pass for these versions. The opposite is seen for versions two and five. All the original tests fail and the accuracy of our predictions is close to 0\%. The results for these four versions can be explained if our model is blindly passing tests and producing a large number of false positives.

\begin{table}[h]
	\caption{Percentage of passed tests in the original fault matrix for the ABS application, compared to the best accuracy obtained in our predictions.}
	\label{tab:abs_1s}
	\centering
{\small 	\begin{tabular}{lrrrrrr}
	\toprule
						& v1 	& v2	& v3 	& v4	& v5 	& overall	\\
	\midrule
	Best Percentage		& 75.73	& 7.87	& 100 	& 100	& 12.27 & 54.93		\\
	Percentage of 1s	& 100	& 0 	& 100	& 100	& 0 	& 54.4		\\
	\bottomrule
	\end{tabular}}
\end{table}

We present, in Table \ref{tab:abs_metrics}, for the best ABS strategy (FREQ),  the values of the described metrics for several training percentages.

\begin{table}[h]
	\caption{Several metrics collected for different training percentages of the best strategy for the ABS application.}
	\label{tab:abs_metrics}
	\centering
{\small 	\begin{tabular}{lrrrr}
	\toprule
 			& 5 		& 10 		&  50		& 100 		\\
\midrule
		ACC	&	54.93 	& 54.4		& 54.4		& 54.4	 	\\
		PPV	&	97.35 	& 97.2		& 97.28		& 97.28		\\
		NPV	&	97.23 	& -			& -			& -			\\
		TPR	&	99.8 	& 100		& 100		& 100		\\
		TNR	&	72.48 	& 0			& 0			& 0			\\
	\bottomrule
	\end{tabular}}
\end{table}

As seen before, the accuracy for this program is always around 54\%. When we use 5\% of the executions as training data,  we achieve very high values of PPV and NPV, which means that our predictions, both positive and negative, are almost always correct. However, while we find most of the positives tests (as shown by TPR), we fail to find a large portion of the negative tests (as shown by the TNR value). As we increase the training percentage, the accuracy, PPV and TPR values remain somewhat the same. TNR drops to 0\% meaning that we don't predict any of the negative tests in the original fault matrix. Also, as the training percentage grows, NPV becomes impossible to calculate. This happens because the denominator is the sum of all negative predictions (both false and true), which becomes 0, as our approach stops making negative predictions after 5\%.

% % % % % % % % % % % % % % % % % % % % % % % % % % % % % %
% GZIP
% % % % % % % % % % % % % % % % % % % % % % % % % % % % % %

It seems that the same is happening with the GZIP application, although the effect isn't as noticeable. Table \ref{tab:gzip_1s} presents, for each of the versions and overall,  the best percentage obtained in the tests (the upper row) and the percentage of positive tests in the original fault matrix (the bottom row).

\begin{table}[h]
	\caption{Percentage of passed tests in the original fault matrix for the GZIP application, compared to the best accuracy obtained in our predictions.}
	\label{tab:gzip_1s}
	\centering
{\small 	\begin{tabular}{lrrrrrr}
	\toprule
						& v1 	& v2	& v3 	& v4	& v5 	& overall	\\
	\midrule
	Best Percentage		& 96.26 & 24.17 & 100 	& 99.53 & 99.07 & 83.18 \\
	Percentage of 1s	& 95.33 & 20.09 & 99.07	& 98.60 & 98.13 & 82.24 \\
	\bottomrule
	\end{tabular}}
\end{table}

These values are always quite similar, although the accuracy achieved with out predictions is higher, which could mean that we can actually find negative tests for the GZIP application, as opposed to what happens with ABS. We can see,  in Table \ref{tab:gzip_metrics}, for the best GZIP strategy,  the values of the described metrics for several training percentages.

\begin{table}[h]
	\caption{Several metrics collected for different training percentages of the best strategy for the GZIP application.}
	\label{tab:gzip_metrics}
	\centering
{\small 	\begin{tabular}{lrrrr}
	\toprule
 			& 5 		& 10 		&  50		& 100 		\\
\midrule
		ACC	&	81.95 & 82.63	& 82.90	& 83.18 	\\
		PPV	&	99.32 & 99.32	& 99.32	& 99.32	\\
		NPV	&	96.00 & 97.90	& 98.82	& 100.00	\\
		TPR	&	99.94 & 99.97	& 99.99	& 100.00	\\
		TNR	&	67.69 & 66.31	& 64.69	& 62.50	\\
	\bottomrule
	\end{tabular}}
\end{table}

The values of PPV and NPV are quite high, meaning that, when we make a predictions (either positive or negative),  we are confident it will be correct. As we increase the training percentage, the number of negative predictions (tn + fn) decreases, but in a good way, as the false negatives decrease more (even reaching 0) than the true negatives. The value of TPR is also quite high and even reaches 100\%, which was expected. When most of our predictions are positive predictions,  we will eventually find all the positive test cases in the original fault matrix. On the other hand, TNR has a rather small value, misclassifying around 40\% of the negative examples (those were the original tests fail). TNR decreases as the training percentage used increases, as the number of positive predictions also increases.

Once again, GZIP appears to have the same false positive problem that ABS shows, but at a smaller scale. Even though we fail to correctly predict some of the negative instances (as highlighted by TNR), when we make a negative prediction, we do so with a high degree of certainty (evidenced by NPV).

\subsection{Overview and Analysis}

% overall
Our approach seems to produce a large number of false positive results. The reason behind this behavior is not clear at this time. One possibility is that the current process of comparing ranges may be too permissive. This is something that should be looked into, especially considering that there is a big difference on the rate of false positives between the two example applications. It may be possible, by analyzing the different characteristics of the two applications, to identify the origin of this problem.

Although there is a difference between the accuracy of the presented strategies, this difference is smaller than expected, especially in cases where the number of monitored variables greatly varies. Generally speaking, the strategies FREQ and COMBAND seem to achieve roughly the same accuracy as others, but manage to do it using a significantly smaller percentage of the training data. This could be an important factor, as it would require a shorter time to perform our fault detection tests.

The first two strategies, which monitor function parameters, assignments to scalar variables and function return values, may have redundant information that causes a degradation in our detection accuracy. One could argue that the return value of a function depends on the assignments that occur inside its body. If this is accepted, then we should only choose to monitor one or the other, but not both. This concept obviously needs a more sophisticated approach.

% abs
The results for the ABS application are mixed, as there are faults that clearly were easier to detect than others. Since the only difference in the versions is the injected faults, we can assume that it is the type of the faults that deteriorates our ability to detect them. One of the injected faults has a direct impact on the output of the program, but does not cause a dramatic change in the learned ranges. This seems to be one weakness of our approach. Another weakness is exposed when a second fault causes the values inside an array to change. We assume this could be detected if we were monitoring ranges of array and pointer variables. However, this is not the case, as none of the tested strategies are prepared to deal with these variables.

% gzip
The results are overall positive for the GZIP application. The only problem arises when one of the injected faults has the ability to drastically change the output of a program without causing major changes on the ranges of the monitored variables. This is exactly what happens in the fault injected in version two, where the compression method is changed when the fault is active.

% summarize the false positive analysis 
Our approach seems to produce a considerable amount of false positive results for both applications tested, although this is more noticeable in the ABS application. Nonetheless, whenever our application makes a negative prediction (stating that the test has failed),  it does so with high accuracy. The problem is that there is still a large number of negative tests, found by traditional comparison methods, that our approach incorrectly labels as positives tests.

We are not able to identify each and every variable whose range is critical to the diagnostic. This was somewhat expected, as the main issue is still to find the critical or collar variables out of all the variables that are used during the execution of an application. Although the instrumentation can be a challenge and efficient techniques are required, the first and most important step, is to find what needs to be instrumented.

Faults that cause and propagate errors and do not violate the learned ranges are not detected. However, these faults can be triggered during execution and our approach is not capable of detecting them. It is possible that, in a given execution context, even if a value is within the learned range, it can cause unpredictable behavior that leads to an erroneous state. For instance, consider the second fault injected in the GZIP application (which had the worst detection accuracy out of all the faults). This fault changes the method used to compress the input data, producing a completely different output. Suppose that a fault causes the value of the variable that holds the compression method to change. Even if we learn the range for this variable (e.g., 1 through 9), such a change can lead to an error state and unexpected output.

These types of control or operation mode variables are difficult to monitor through ranges, as they are inherently discrete, meaning that ranges may be an unsuitable representation in the first place. In addition to this problem, we still need to ensure that we exercise all of the operation modes during the training phase. For these types of variables an approach using only ranges does not seem sufficient. Nevertheless, this specific example, in which the compression method is changed, looks quite difficult to detect even if other approaches were used. Here, even a simple change of the value of a single variable can lead to a much different program output, even if the value is still within a well.defined range.

Our approach seems to have issues dealing with faults that drastically change the control flow of the application as value ranges may not be the most suitable representation to deal with these types of faults. However, this approach shows potential and we manage to achieve interesting and motivating results in the GZIP application. We expected to find some difficulties when using this approach by itself. It becomes obvious that it needs some refinement and that it would greatly benefit to be supported and complemented by other approaches, such as approaches that take control flow into account.

\section{Summary}

In this chapter we presented the results of our experimental evaluation using two applications. We compared the ability of our approach detecting faults to the more traditional method of comparing the outputs of the executions. We used two applications with different characteristics. The first is the GZIP compressor and the other is an ABS simulation.

The results for the ABS application are mixed as we only achieve a top accuracy of around 54\% overall when considering all the tested faults. It is clear that some faults are easier to detect than others, as we achieve perfect or near-perfect in some of the faults but have accuracies of around 13\% in others. One key feature that is missing from our strategies is array monitoring as some of the faults seem to have a large impact on array values which we are not able to detect. Monitoring arrays will, in principle, improve the achieved results. However, it maybe difficult to implement this as we need a suitable strategy. For instance, in most cases, it would not be feasible to monitor every single position of the array. One possibility would be to maintain a range for the entire array, but this may not be accurate.

The results are better for the GZIP application as we manage to achieve an overall accuracy of around 83\%. The worst result for a single fault of GZIP highlights another key feature that is missing in our approach, which is the ability to reliably monitor control variables. These commonly assume only a small set of discrete values which are not well captured by our concept of range. In order to overcome this limitation, we would need to possibly use our approach to complement others that may be better suited to deal with this class fo variables. 

% JPs & slowdowns
Performing the analyses and transformations needed for our approach would prove to be massive tasks if performed manually. Moreover it would more likely result in user-induced faults as the process would be quite long and tedious. Our framework manages to analyze and act over a large number of program critical points in a short amount of time. For ABS, we analyze at most 1742 points and act over 129. These numbers become 5466 and 483 for the GZIP application. This is performed in only 5 and 7 seconds for ABS and GZIP respectively.

Finally, the instrumentation which is key to this approach results in slowdowns of around 0.86 for ABS and 0.73 for GZIP. With a more sophisticated and careful choice of variables, this technique could perhaps be used for online fault detection that could trigger fault-recovery policies.

% % % % % % % % % % % % % % % % % % % %
% CONCLUSIONS

\chapter{Conclusions}\label{chap:conc}

% description of the goals
We started our work with the goal of developing a compiler capable of applying custom code instrumentation strategies that could be used to detect the presence of faults within an application. We extended an existing source-to-source compiler with analyses and transformations passes, as well as support for the LARA language. This provides us with and Aspect-Oriented Programming approach that separates the original application and the definition of the instrumentation strategies.

% what our compiler can do
A source-to-source compiler extended with LARA language support provides great flexibility, as it can be used for a multitude of different tasks such as source code analysis, optimizations and instrumentation. There is a large number of possibilities for instrumentation as the compiler exposes fine-grain selection of code structures that users can use to target very specific and critical code sections. Furthermore, because this approach forces the separation of the original application and the specification of strategies, we can rapidly test several different instrumentation possibilities without heavy changes to the application.

% our approach using the ranges
Although the compiler can be used for several different tasks and instrumentation strategies (e.g., capturing control flow signatures), our focus was code instrumentation that could provide us with the range of the values taken by specific, critical variables. During an initial training phase, we execute our applications and try to exercise as many modules and features as possible. As a result, we learn, for each of the monitored variables a range of values that were taken. Then, in the testing phase, when a value is assigned to a variable, we check if it falls outside the learned range. When this happens we consider that a fault was detected.

% overview of the ABS results
We evaluated our approach on two different applications, the GZIP compression utility and an ABS simulator provided by our industrial partner. We tested different strategies which focused on choosing distinct sets of variables to monitor. The results for the ABS application are mixed. We achieved an overall accuracy of around 54\% for he 5 different faults tested. For two of the versions we achieve perfect accuracy while on two others, we don't go beyond 13\%. As seen, some of the faults are easier to detected than others. For instance, if the fault leads to a change in the values of an array, we are not able to pinpoint it on the array and rely on monitoring other variables, which may depend on the array. This is one of the directions for future work, as at the present, there is no reliable way to test the range of array variables.

% overview of the GZIP results
As for the GZIP application, the results are positive and we obtain an overall accuracy of around 83\% for the 5 different injected faults. With the exception of a single fault, all the tested strategies produce good results with accuracies that are either perfect or close ot perfect. The only negative point is the second fault, in which our approach only achieves around 22\% accuracy. This seems to be one of the weaknesses of this approach, capturing the meaning of control variables, which is difficult to do with just a value range.

% instrumentation points and instrumentation time
The number of source code points that are analyzed and instrumented means that a manual approach is impractical, as it will require a big effort by the user and will, almost surely, results in a great time loss and possible errors introduced in the code. In the ABS application, our strategies analyzed a maximum of 1742 points in the code and acted over 129 of those. As for GZIP, our strategies analyzed a maximum of 5466 code locations and inserted code at 483 of those. Clearly, this is an effort that is better suited for an automatic approach such as ours, especially considering that these strategies execute much faster than any user could if performing these tasks. The instrumentation process takes at most 5 seconds for ABS and 7 seconds for GZIP. There is the obvious overhead of learning the LARA language and developing the strategies, but the benefits still outweigh the disadvantages. Moreover, these overheads will be reduced as more and more strategies are developed. It is also worth noting that the strategies developed with the LARA language can be applied to other applications.

% slowdown and the possibility of using this as an online measure
In the worst case, for the ABS application, there is a slowdown of 0.86. With GZIP, the worst strategy causes a slowdown 0.73. These slowdowns account only for the execution time of the binaries instrumented to collect the range information. Although there would still be the overhead of testing the ranges to detect a fault, we believe it would be possible to use this approach as an online measure for fault detection that could trigger proper fault recovery mechanisms. In order to do this, we need more sophisticated strategies that can select only the most critical variables in the code. This would reduce the number of online tests which, in turn, would provide better slowdowns.

% the results are not great but our infrastructure can be used to perform this analysis
Although the results show some instances where our approach mispredicts a considerable amount of tests, it is clear that our infrastructure is able to implement custom instrumentation strategies capable of providing fault detection diagnosis. In the future, new and more sophisticated strategies can be implemented. These should enable better diagnostic accuracy for fault detection and location. Because of this, we believe that there is clear room for improvement within our approach.

% mentioning one of the most noticeable drawbacks
One of the negative features that stood out from the analysis on the experimental results, is that it is quite difficult to detect faults that target control variables using just our approach. These variables are responsible to direct the flow of the execution and to choose different operation modes. In practice, these variables can take a value that is well within the learned range, but still produce an output that is largely different from what is expected.
% this was expected, how can we improve?, possible future study
We expected our approach to not detect every fault as well as a manual approach, as it is quite difficult to monitor every variable of interest in the application. We believe that one way to improve the diagnostic accuracy is to complement our approach with different, already-proven techniques, such as a control flow analysis that could help with control variables. Another interesting future study, would focus on finding which types of faults are detected by our approach but not by others more commonly used. The support provided by MANET and LARA for specifying monitoring strategies will be of great help in further experiments.

\bibliography{autoseerreport}
\bibliographystyle{plain}

\appendix

\lstset{ %
  language=C,
  backgroundcolor=\color{white},   % choose the background color; you must add \usepackage{color} or \usepackage{xcolor}
  basicstyle=\scriptsize\ttfamily,        % the size of the fonts that are used for the code
  breaklines=true,                 % sets automatic line breaking
  deletekeywords={...},            % if you want to delete keywords from the given language
  extendedchars=true,              % lets you use non-ASCII characters; for 8-bits encodings only, does not work with UTF-8
  keywordstyle=\bfseries\color{blue},       % keyword style
  morekeywords={aspectdef, call, begin, end, input, output, select, apply, insert},            % if you want to add more keywords to the set
  numbers=left,                    % where to put the line-numbers; possible values are (none, left, right)
  numbersep=5pt,                   % how far the line-numbers are from the code
  numberstyle=\tiny\color{gray}, % the style that is used for the line-numbers
  showtabs=false,                  % show tabs within strings adding particular underscores
  tabsize=2,                       % sets default tabsize to 2 spaces
  stringstyle=\color{red},     % string literal style
  showstringspaces=false,          % underline spaces within strings only
}

\chapter{LARA Strategies Code}\label{app:LARA}

\section*{ASCV3}

\begin{lstlisting}[boxpos=b]
aspectdef AutoSeerCurrentv3
	
	input
		onlyPrimitives=true;
	end
	
	var laraObj = new LaraObject();

	call PrepareProgram();
	call InstrumentParameters(laraObj, onlyPrimitives);
	call InstrumentWrites(laraObj, onlyPrimitives);
	call InstrumentReturns(laraObj, onlyPrimitives);
	call SetupInstrumentation(laraObj);
end

aspectdef PrepareProgram
	
	select program end
	apply
		$program.SingleDeclarator();
		$program.AssignmentExpansion();
		$program.UnaryExpansion();
		$program.NormalizeReturn();
	end
end

aspectdef SetupInstrumentation

	input
		laraObject
	end

	var size = laraObject.getTotal();
	
	select function end
	apply
		insert before '#include "range.h"';
		insert before 'Range __r_a[ [[size]] ];';
		break;
	end
	
	select function{"main"}.first_stmt end
	apply
		insert before 'range_init(__r_a, [[size]]);\natexit(print_ranges_to_file);';
	end

	var PrintFunctionJS = function (laraObject) {

		var code = '\nFILE* file;\ndouble inf = 1.0 / 0.0;\n';
		code += 'file = fopen("./output/range.mon", "w");\n\n';

		code += 'if(file == NULL){printf("Could not open the output file.\\n"); return;}\n\n';
		
		code += 'fprintf(file, "' + size + '\\n");\n';
		
		for(var type in laraObject) {
			
			if(!laraObject.hasOwnProperty(type) || typeof laraObject[type] == 'function') {
				continue;
			}
			
			code += 'fprintf(file, "t#' + type + '\\n");\n';
			
			for(var func in laraObject[type]){
			
				code += 'fprintf(file, "f#' + func + '\\n");\n';
				
				for (var vari in laraObject[type][func]) {
				
					code += 'if(__r_a[' + laraObject[type][func][vari] + '].min != 1./0.) {\n';
					code += 'fprintf(file, "v#' + vari + '#' + laraObject[type][func][vari] + '");\n';
					
					code += 'fprintf(file, "#%.2f#%.2f\\n", __r_a[' + laraObject[type][func][vari] + '].min, __r_a[' + laraObject[type][func][vari] + '].max);\n}\n';
					
				}
			}
		}

		code += '\nfclose(file);\n';

		return code;
	};
	var code = PrintFunctionJS(laraObject);
	
	select function{"main"} end
	apply
		insert before 'void print_ranges_to_file() {\n[[code]]}';
	end
end

aspectdef InstrumentParameters
	
	input
		laraObject,
		onlyPrimitives
	end
	
	P: select function.param end
	F: select function.body.first_stmt end
	apply to F::P
	
		if(onlyPrimitives)
			if(!$param.is_primitive)
				continue;
				
		var index = laraObject.getId('param', $function.name, $param.name);
		$first_stmt.insert before 'range_update(&(__r_a[ [[index]] ]), [[$param.name]]); /* param */';
	end
	condition
		$function.name != "main" &&
		!$param.is_array &&
		!$param.is_pointer &&
		!$param.is_struct &&
		$param.name != ""
	end
end

aspectdef InstrumentWrites
	
	input
		laraObject,
		onlyPrimitives
	end
	
	select file.function.var end
	apply
		if(onlyPrimitives)
			if(!$var.is_primitive)
				continue;
	
		var index = laraObject.getId('write', $function.name, $var.name).toString();
		$var.insert after 'range_update(&(__r_a[ [[index]] ]), [[$var.name]]);';
	end
	condition
		!$var.in_loop_header &&
		$var.reference == "write" &&
		!$var.is_array &&
		!$var.is_pointer &&
		!$var.is_struct 
	end
end

aspectdef InstrumentReturns

	input
		laraObject,
		onlyPrimitives
	end

	select function.return.var end
	apply
		if(onlyPrimitives)
			if(!$var.is_primitive)
				continue;
				
		var index = laraObject.getId('return', $function.name, $var.name).toString();
		$return.insert before 'range_update(&(__r_a[ [[index]] ]), [[$var.name]]);';
	end
	condition
		!$var.is_array &&
		!$var.is_pointer &&
		!$var.is_struct 
	end
end
\end{lstlisting}

\section*{ASCV3\_s}

\begin{lstlisting}[boxpos=b]
aspectdef AutoSeerCurrentv3_structs
	
	input
		onlyPrimitives=true;
	end
	
	var laraObj = new LaraObject();
	
	call PrepareProgram();
	call InstrumentParameters(laraObj, onlyPrimitives);
	call InstrumentWrites(laraObj, onlyPrimitives);
	call InstrumentReturns(laraObj, onlyPrimitives);
	call SetupInstrumentation(laraObj);
end



aspectdef PrepareProgram
	
	select program end
	apply
		$program.perform SingleDeclarator();
		$program.perform AssignmentExpansion();
		$program.perform StructAssignmentDecomposition();
		$program.perform UnaryExpansion();
		$program.perform NormalizeReturn();
	end
end

aspectdef SetupInstrumentation

	input
		laraObject
	end

	var size = laraObject.getTotal();
	
	select function end
	apply
		insert before '#include "range.h"';
		insert before 'Range __r_a[ [[size]] ];';
		break;
	end
	
	select function{"main"}.first_stmt end
	apply
		insert before 'range_init(__r_a, [[size]]);\natexit(print_ranges_to_file);';
	end

	var PrintFunctionJS = function (laraObject) {

		var code = '\nFILE* file;\ndouble inf = 1.0 / 0.0;\n';
		code += 'file = fopen("./output/range.mon", "w");\n\n';

		code += 'if(file == NULL){printf("Could not open the output file.\\n"); return;}\n\n';

		code += 'fprintf(file, "' + size + '\\n");\n';
		
		for(var type in laraObject) {
			
			if(!laraObject.hasOwnProperty(type) || typeof laraObject[type] == 'function') {
				continue;
			}
			
			code += 'fprintf(file, "t#' + type + '\\n");\n';
			
			for(var func in laraObject[type]){
			
				code += 'fprintf(file, "f#' + func + '\\n");\n';
				
				for (var vari in laraObject[type][func]) {
				
					code += 'if(__r_a[' + laraObject[type][func][vari] + '].min != 1./0.) {\n';
					code += 'fprintf(file, "v#' + vari + '#' + laraObject[type][func][vari] + '");\n';
					
					code += 'fprintf(file, "#%.2f#%.2f\\n", __r_a[' + laraObject[type][func][vari] + '].min, __r_a[' + laraObject[type][func][vari] + '].max);\n}\n';
					
				}
			}
		}

		code += '\nfclose(file);\n';

		return code;
	};
	var code = PrintFunctionJS(laraObject);
	
	select function{"main"} end
	apply
		insert before 'void print_ranges_to_file() {\n[[code]]}';
	end
end

aspectdef InstrumentParameters
	
	input
		laraObject,
		onlyPrimitives
	end
	
	P: select function.param end
	F: select function.body.first_stmt end
	apply to F::P
	
		if(onlyPrimitives)
			if(!$param.is_primitive)
				continue;
				
		var index = laraObject.getId('param', $function.name, $param.name);
		$first_stmt.insert before 'range_update(&(__r_a[ [[index]] ]), [[$param.name]]); /* param */';
	end
	condition
		$function.name != "main" &&
		!$param.is_array &&
		!$param.is_pointer &&
		!$param.is_struct &&
		$param.name != ""
	end
end

aspectdef InstrumentWrites
	
	input
		laraObject,
		onlyPrimitives
	end


	select file.function.var{reference == 'write'} end
	apply

		if(onlyPrimitives)
			if(!$var.is_primitive)
				continue;
		var index = laraObject.getId('write', $function.name, $var.name);
		$var.insert after 'range_update(&(__r_a[ [[index]] ]), [[$var.name]]);';
		
	end
	condition
		!$var.in_loop_header &&
		!$var.is_array &&
		!$var.is_struct &&
		!$var.is_pointer
	end
end

aspectdef InstrumentReturns

	input
		laraObject,
		onlyPrimitives
	end

	select function.return.var end
	apply
		if(onlyPrimitives)
			if(!$var.is_primitive)
				continue;
	
		var index = laraObject.getId('return', $function.name, $var.name).toString();
		$return.insert before 'range_update(&(__r_a[ [[index]] ]), [[$var.name]]);';
	end
	condition
		!$var.is_array &&
		!$var.is_pointer &&
		!$var.is_struct 
	end
end
\end{lstlisting}

\section*{FREQ}

\begin{lstlisting}[boxpos=b]
aspectdef Frequency
	
	input
		reportName='',
		onlyPrimitives=true
	end
	
	var laraObj = new LaraObject();
	
	call res : ChooseVarsToMonitor(0.01, reportName);
	var varsToMonitor = res.varsToMonitor;
	
	call PrepareProgram();
	call InstrumentWrites(laraObj, varsToMonitor, onlyPrimitives);
	call SetupInstrumentation(laraObj);
end

aspectdef ChooseVarsToMonitor

	input threshold, reportName end
	output varsToMonitor end
	varsToMonitor = [];

	var reportPath = './reports/' + reportName + 'freq.report';
	
	var report = fileToJSON(reportPath);
	var total = report['total'];
	for(funcVar in report) {

		var ratio = report[funcVar] / total;
		if(ratio > threshold) {

			varsToMonitor.push(funcVar);
		}
	}
	varsToMonitor.pop();
end

aspectdef PrepareProgram
	
	select program end
	apply
		$program.perform SingleDeclarator();
		$program.perform AssignmentExpansion();
		$program.perform UnaryExpansion();
		$program.perform NormalizeReturn();
	end
end

aspectdef InstrumentWrites
	
	input
		laraObject,
		varsToMonitor,
		onlyPrimitives
	end
	
	var present = function(array, key) {
		
		for(el in array) {
			
			if(array[el] === key) {
				return true;
			}
		}
		return false;
	};
	
	select function.var end
	apply
	
		if(onlyPrimitives)
			if(!$var.is_primitive)
				continue;
	
		var key = $function.name + '#' + $var.name;
		if(present(varsToMonitor, key)) {
			
			var index = laraObject.getId('write', $function.name, $var.name);
			$var.insert after 'range_update(&(__r_a[ [[index]] ]), [[$var.name]]);';
		}
	end
	condition
		!$var.in_loop_header &&
		$var.reference == "write" &&
		!$var.is_array &&
		!$var.is_pointer &&
		!$var.is_struct 
	end
end

aspectdef SetupInstrumentation

	input
		laraObject
	end

	var size = laraObject.getTotal();
	
	select function end
	apply
		insert before '#include "range.h"';
		insert before 'Range __r_a[ [[size]] ];';
		break;
	end
	
	select function{"main"}.first_stmt end
	apply
		insert before 'range_init(__r_a, [[size]]);\natexit(print_ranges_to_file);';
	end

	var PrintFunctionJS = function (laraObject) {

		var code = '\nFILE* file;\ndouble inf = 1.0 / 0.0;\n';
		code += 'file = fopen("./output/range.mon", "w");\n\n';

		code += 'if(file == NULL){printf("Could not open the output file.\\n"); return;}\n\n';
		
		code += 'fprintf(file, "' + size + '\\n");\n';
		
		for(var type in laraObject) {
			
			if(!laraObject.hasOwnProperty(type) || typeof laraObject[type] == 'function') {
				continue;
			}
			
			code += 'fprintf(file, "t#' + type + '\\n");\n';
			
			for(var func in laraObject[type]){
			
				code += 'fprintf(file, "f#' + func + '\\n");\n';
				
				for (var vari in laraObject[type][func]) {
				
					code += 'if(__r_a[' + laraObject[type][func][vari] + '].min != 1./0.) {\n';
					code += 'fprintf(file, "v#' + vari + '#' + laraObject[type][func][vari] + '");\n';

					code += 'fprintf(file, "#%.2f#%.2f\\n", __r_a[' + laraObject[type][func][vari] + '].min, __r_a[' + laraObject[type][func][vari] + '].max);\n}\n';
					
				}
			}
		}

		code += '\nfclose(file);\n';

		return code;
	};
	var code = PrintFunctionJS(laraObject);
	
	select function{"main"} end
	apply
		insert before 'void print_ranges_to_file() {\n[[code]]}';
	end
end
\end{lstlisting}

\section*{FANIN}

\begin{lstlisting}[boxpos=b]
aspectdef FanIn
	
	input
		reportName='',
		onlyPrimitives=true
	end
	
	var laraObj = new LaraObject();
	
	call res : ChooseVarsToMonitor(2, reportName);
	var varsToMonitor = res.varsToMonitor;
	
	call PrepareProgram();
	call InstrumentWrites(laraObj, varsToMonitor, onlyPrimitives);
	call SetupInstrumentation(laraObj);
end

aspectdef ChooseVarsToMonitor

	input threshold, reportName end
	output varsToMonitor end
	varsToMonitor = [];

	var reportPath = './reports/' + reportName + 'fan_in.report';

	var report = fileToJSON(reportPath);
	for(funcVar in report) {

		if(report[funcVar] > threshold) {

			varsToMonitor.push(funcVar);
		}
	}
end

aspectdef PrepareProgram
	
	select program end
	apply
		$program.perform SingleDeclarator();
		$program.perform AssignmentExpansion();
		$program.perform UnaryExpansion();
		$program.perform NormalizeReturn();
	end
end

aspectdef InstrumentWrites
	
	input
		laraObject,
		varsToMonitor,
		onlyPrimitives
	end
	
	var present = function(array, key) {
		
		for(el in array) {
			
			if(array[el] === key) {
				return true;
			}
		}
		return false;
	};
	
	select function.var end
	apply
	
		if(onlyPrimitives)
			if(!$var.is_primitive)
				continue;
		
		var key = $function.name + '#' + $var.name;
		if(present(varsToMonitor, key)) {
			
			var index = laraObject.getId('write', $function.name, $var.name);
			$var.insert after 'range_update(&(__r_a[ [[index]] ]), [[$var.name]]);';
		}
	end
	condition
		!$var.in_loop_header &&
		$var.reference == "write" &&
		!$var.is_array &&
		!$var.is_pointer &&
		!$var.is_struct 
	end
end

aspectdef SetupInstrumentation

	input
		laraObject
	end

	var size = laraObject.getTotal();
	
	select function end
	apply
		insert before '#include "range.h"';
		insert before 'Range __r_a[ [[size]] ];';
		break;
	end
	
	select function{"main"}.first_stmt end
	apply
		insert before 'range_init(__r_a, [[size]]);\natexit(print_ranges_to_file);';
	end

	var PrintFunctionJS = function (laraObject) {

		var code = '\nFILE* file;\ndouble inf = 1.0 / 0.0;\n';
		code += 'file = fopen("./output/range.mon", "w");\n\n';

		code += 'if(file == NULL){printf("Could not open the output file.\\n"); return;}\n\n';

		code += 'fprintf(file, "' + size + '\\n");\n';
		
		for(var type in laraObject) {

			if(!laraObject.hasOwnProperty(type) || typeof laraObject[type] == 'function') {
				continue;
			}
			
			code += 'fprintf(file, "t#' + type + '\\n");\n';
			
			for(var func in laraObject[type]){
			
				code += 'fprintf(file, "f#' + func + '\\n");\n';
				
				for (var vari in laraObject[type][func]) {
				
					code += 'if(__r_a[' + laraObject[type][func][vari] + '].min != 1./0.) {\n';
					code += 'fprintf(file, "v#' + vari + '#' + laraObject[type][func][vari] + '");\n';

					code += 'fprintf(file, "#%.2f#%.2f\\n", __r_a[' + laraObject[type][func][vari] + '].min, __r_a[' + laraObject[type][func][vari] + '].max);\n}\n';
					
				}
			}
		}

		code += '\nfclose(file);\n';

		return code;
	};
	var code = PrintFunctionJS(laraObject);
	
	select function{"main"} end
	apply
		insert before 'void print_ranges_to_file() {\n[[code]]}';
	end
end
\end{lstlisting}

\section*{COMBAND}

\begin{lstlisting}[boxpos=b]
aspectdef CombineAnd
	
	input
		reportName='',
		onlyPrimitives=true
	end
	
	var laraObj = new LaraObject();
	
	call res : ChooseVarsToMonitor(2, 0.01, reportName);
	var varsToMonitor = res.varsToMonitor;

	call PrepareProgram();
	call InstrumentWrites(laraObj, varsToMonitor, onlyPrimitives);
	call SetupInstrumentation(laraObj);
end

aspectdef ChooseVarsToMonitor

	input
		fanInThreshold,
		freqThreshold,
		reportName
	end
	output varsToMonitor end
	varsToMonitor = [];

	var fanInReport = fileToJSON('./reports/' + reportName + 'fan_in.report');
	var freqReport = fileToJSON('./reports/' + reportName + 'freq.report');

	var isFreqGood = function (value, total, threshold) {
		return	(value != undefined) &&
				(value / total > threshold);
	};
	
	var isFanInGood = function (value, threshold) {
		return	(value != undefined) &&
				(value > threshold);
	};

	var total = freqReport['total'];
	
	for(funcVar in fanInReport)
		if(isFanInGood(fanInReport[funcVar], fanInThreshold)) 
			if(isFreqGood(freqReport[funcVar], total, freqThreshold)) 
				varsToMonitor.push(funcVar);
end

aspectdef PrepareProgram
	
	select program end
	apply
		$program.perform SingleDeclarator();
		$program.perform AssignmentExpansion();
		$program.perform UnaryExpansion();
		$program.perform NormalizeReturn();
	end
end

aspectdef InstrumentWrites
	
	input
		laraObject,
		varsToMonitor,
		onlyPrimitives
	end
	
	var present = function(array, key) {
		
		for(el in array) {
			
			if(array[el] === key) {
				return true;
			}
		}
		return false;
	};
	
	select function.var end
	apply
	
		if(onlyPrimitives)
			if(!$var.is_primitive)
				continue;
	
		var key = $function.name + '#' + $var.name;
		if(present(varsToMonitor, key)) {
			
			var index = laraObject.getId('write', $function.name, $var.name);
			$var.insert after 'range_update(&(__r_a[ [[index]] ]), [[$var.name]]);';
		}
	end
	condition
		!$var.in_loop_header &&
		$var.reference == "write" &&
		!$var.is_array &&
		!$var.is_pointer &&
		!$var.is_struct 
	end
end

aspectdef SetupInstrumentation

	input
		laraObject
	end

	var size = laraObject.getTotal();
	
	select function end
	apply
		insert before '#include "range.h"';
		insert before 'Range __r_a[ [[size]] ];';
		break;
	end
	
	select function{"main"}.first_stmt end
	apply
		insert before 'range_init(__r_a, [[size]]);\natexit(print_ranges_to_file);';
	end

	var PrintFunctionJS = function (laraObject) {

		var code = '\nFILE* file;\ndouble inf = 1.0 / 0.0;\n';
		code += 'file = fopen("./output/range.mon", "w");\n\n';

		code += 'if(file == NULL){printf("Could not open the output file.\\n"); return;}\n\n';
		
		code += 'fprintf(file, "' + size + '\\n");\n';
		
		for(var type in laraObject) {
			
			if(!laraObject.hasOwnProperty(type) || typeof laraObject[type] == 'function') {
				continue;
			}
			
			code += 'fprintf(file, "t#' + type + '\\n");\n';
			
			for(var func in laraObject[type]){
			
				code += 'fprintf(file, "f#' + func + '\\n");\n';
				
				for (var vari in laraObject[type][func]) {
				
					code += 'if(__r_a[' + laraObject[type][func][vari] + '].min != 1./0.) {\n';
					code += 'fprintf(file, "v#' + vari + '#' + laraObject[type][func][vari] + '");\n';
					
					code += 'fprintf(file, "#%.2f#%.2f\\n", __r_a[' + laraObject[type][func][vari] + '].min, __r_a[' + laraObject[type][func][vari] + '].max);\n}\n';
					
				}
			}
		}

		code += '\nfclose(file);\n';

		return code;
	};
	var code = PrintFunctionJS(laraObject);
	
	select function{"main"} end
	apply
		insert before 'void print_ranges_to_file() {\n[[code]]}';
	end
end
\end{lstlisting}

\section*{COMBOR}

\begin{lstlisting}[boxpos=b]
aspectdef CombineOr

	input
		reportName='',
		onlyPrimitives=true
	end
	
	var laraObj = new LaraObject();

	call res : ChooseVarsToMonitor(2, 0.01, reportName);
	var varsToMonitor = res.varsToMonitor;

	call PrepareProgram();
	call InstrumentWrites(laraObj, varsToMonitor, onlyPrimitives);
	call SetupInstrumentation(laraObj);
end

aspectdef ChooseVarsToMonitor

	input
		fanInThreshold,
		freqThreshold,
		reportName
	end
	output varsToMonitor end
	varsToMonitor = [];

	var tempSet = {};

	var fanInReport = fileToJSON('./reports/' + reportName + 'fan_in.report');
	var freqReport = fileToJSON('./reports/' + reportName + 'freq.report');

	for(funcVar in fanInReport) {

		if(fanInReport[funcVar] > fanInThreshold) {

			tempSet[funcVar] = 1;
		}
	}

	var total = freqReport['total'];
	for(funcVar in freqReport) {

		var ratio = freqReport[funcVar] / total;
		if(ratio > freqThreshold) {

			tempSet[funcVar] = 1;
		}
	}
	tempSet['total'] = undefined;
	
	for(funcVar in tempSet) {

		if(tempSet[funcVar] == 1)
			varsToMonitor.push(funcVar);
	}
end

aspectdef PrepareProgram
	
	select program end
	apply
		$program.perform SingleDeclarator();
		$program.perform AssignmentExpansion();
		$program.perform UnaryExpansion();
		$program.perform NormalizeReturn();
	end
end

aspectdef InstrumentWrites
	
	input
		laraObject,
		varsToMonitor,
		onlyPrimitives
	end
	
	var present = function(array, key) {
		
		for(el in array) {
			
			if(array[el] === key) {
				return true;
			}
		}
		return false;
	};
	
	select function.var end
	apply
	
		if(onlyPrimitives)
			if(!$var.is_primitive)
				continue;
	
		var key = $function.name + '#' + $var.name;
		if(present(varsToMonitor, key)) {
			
			var index = laraObject.getId('write', $function.name, $var.name);
			$var.insert after 'range_update(&(__r_a[ [[index]] ]), [[$var.name]]);';
		}
	end
	condition
		!$var.in_loop_header &&
		$var.reference == "write" &&
		!$var.is_array &&
		!$var.is_pointer &&
		!$var.is_struct 
	end
end

aspectdef SetupInstrumentation

	input
		laraObject
	end

	var size = laraObject.getTotal();
	
	select function end
	apply
		insert before '#include "range.h"';
		insert before 'Range __r_a[ [[size]] ];';
		break;
	end
	
	select function{"main"}.first_stmt end
	apply
		insert before 'range_init(__r_a, [[size]]);\natexit(print_ranges_to_file);';
	end

	var PrintFunctionJS = function (laraObject) {

		var code = '\nFILE* file;\ndouble inf = 1.0 / 0.0;\n';
		code += 'file = fopen("./output/range.mon", "w");\n\n';

		code += 'if(file == NULL){printf("Could not open the output file.\\n"); return;}\n\n';

		code += 'fprintf(file, "' + size + '\\n");\n';
		
		for(var type in laraObject) {
			
			if(!laraObject.hasOwnProperty(type) || typeof laraObject[type] == 'function') {
				continue;
			}
			
			code += 'fprintf(file, "t#' + type + '\\n");\n';
			
			for(var func in laraObject[type]){
			
				code += 'fprintf(file, "f#' + func + '\\n");\n';
				
				for (var vari in laraObject[type][func]) {
				
					code += 'if(__r_a[' + laraObject[type][func][vari] + '].min != 1./0.) {\n';
					code += 'fprintf(file, "v#' + vari + '#' + laraObject[type][func][vari] + '");\n';

					code += 'fprintf(file, "#%.2f#%.2f\\n", __r_a[' + laraObject[type][func][vari] + '].min, __r_a[' + laraObject[type][func][vari] + '].max);\n}\n';
					
				}
			}
		}

		code += '\nfclose(file);\n';

		return code;
	};
	var code = PrintFunctionJS(laraObject);
	
	select function{"main"} end
	apply
		insert before 'void print_ranges_to_file() {\n[[code]]}';
	end
end
\end{lstlisting}

\end{document}